\documentclass[aps,fleqn,preprint]{revtex4}

\usepackage{amsmath}
\usepackage{graphicx}
\usepackage{epsfig}
\usepackage{subfigure}
\usepackage{amsfonts}
\usepackage{amssymb}
\usepackage{amsthm}
\usepackage{newlfont}
\usepackage{epstopdf}

\usepackage[section] {placeins}
\usepackage{morefloats}

\bibliography{thesis_ref.bib}

\begin{document}

\title{Floquet calculation of harmonic generation from hydrogen molecular ions
in strong monochromatic laser fields}

\author{Ts.~Tsogbayar}
\author{M.~Horbatsch}
\affiliation{Department of Physics and Astronomy, York University, 4700 Keele Street, Toronto, Ontario, Canada M3J 1P3}

\begin{abstract}

We present Floquet calculations of high harmonic generation (HHG) for the lowest two electronic states of the 
$\mbox{H}_2^+$ ion by strong continuous-wave laser fields. We solve the non-Hermitian matrix problem to get 
accurate solutions to the periodic time-dependent Schr\"odinger equation (TDSE) by applying a pseudospectral representation combined with
a complex absorbing potential method. This represents an alternative approach to direct TDSE solutions to obtain the harmonic spectra for the 
ion. We compare our HHG rates for the lower and upper states of $\mbox{H}^{+}_{2}$, which correspond
to the gerade and ungerade ground states in the field-free case, with previously obtained results in the literature.     
We show that the enhancement of the ionization rates at the critical internuclear separation $R_{c}\approx 8\,au$ plays some
role in the appearance of very strong harmonic orders $n=5-11$ at $\lambda = 1064\,nm$ and $n=5-9$ at $\lambda = 800\,nm$ and 
intensity $I=10^{14}\,W/cm^{2}$. 

\end{abstract}

\maketitle

\section{Introduction}

From both theoretical and experimental points of view, high-harmonic generation (HHG) is one of the most studied nonlinear phenomena for atoms and molecules interacting with an intense laser field, in which the system emits radiation at multiples of the laser frequency \cite{Piraux}.  The physical mechanism of HHG is well understood for atoms using a three-step model \cite{Kul,Cor,Lew}: (i) the electron is released by tunnel ionization from the atom core; (ii) the free electron is accelerated by the oscillating laser field, and later is driven back to the core; (iii) the electron can recombine with the core to emit a high-energy photon. This semiclassical formulation for the three-step-model is based on the strong-field approximation (SFA) by Lewenstein et al \cite{Lew}. The model predicts a plateau in the harmonic spectra where many harmonics have similar strength, and it ends with a sharp cutoff. At the cutoff the maximum energy is well approximated by the simple and universal formula $I_{p} + 3.17 U_{p}$, where $I_{p}$ is the ionization potential of the atom and $U_{p}$ is the pondermotive potential, defined as $U_{p} = (F/2\omega)^2$, with $F$ the laser electric field strength and $\omega$, the angular frequency in atomic units, respectively.  The cutoff position can be estimated by
\begin{equation}\label{cutoffn}
   N_{max} = (I_{p} + 3.17 U_{p}) / \omega. 
\end{equation}

For symmetric diatomic molecules Kopold et al \cite{Kop} extended the discussion of a (semi)-classical
cutoff formula. They investigated two phenomena, which can become particularly important if one
considers dissociating molecules, i.e., systems at large internuclear separation $R$. The so-called 
simpleman formula (1) is modified, since the ionized electron produced at nucleus $A$ upon
re-collision can be re-combining either at nucleus $A$ or $B$. This can lead to a cutoff that
is higher than the atomic one given in Eq. (1). In addition, there is the possibility
that the field ionizes an electron at atom $A$, accelerates it, and recombination occurs directly
at atom $B$. These classical cutoff positions have to be taken with a grain of salt, since they
ignore the potential role of the Compton profile of the initial state, and the argumentation
based on electron localization during ionization and recombination makes sense only at
large $R$, if the molecular orbital nature of the states is taken into account. Nevertheless,
Ref.\cite{Kop} serves to illustrate that the cutoff energies
represent stationary points at which enhanced HHG should be observed. Evidence is presented in \cite{Kop}
from quantum calculations in a zero-range model potential for molecular cutoffs higher than 
Eq.(1), increasing the coefficient from 3.17 by up to a third for the cases considered.

Theoretical investigations for diatomic molecules,  such as, the $\mbox{H}_{2}$ molecule and the $\mbox{H}^{+}_{2}$ ion had been initially carried out by Krause et al \cite{Krause} and Zuo et al \cite{Zuo1, Zuo2}. They  performed a direct numerical solution of the TDSE to obtain the HHG spectra. An alternative approach is the Floquet formalism which was employed successfully by Potvliege and  Shakeshaft \cite{ShakePot} to obtain the HHG spectra for $\mbox{H}$ atoms using Sturmian basis functions. A treatment of HHG for complex atoms in intense laser fields based on $R$-matrix-Floquet theory has been given by Burke et al \cite{Burke,Burke2}. Yet another method for time-periodic systems is the Floquet approach combined with complex rotation of the coordinate \cite{Ben,Ben2}.  

For atomic hydrogen, the hydrogen molecule and molecular ion calculations of HHG spectra within the Floquet method combined with a complex rotated coordinate have been extensively investigated by S-I.~Chu and his co-workers \cite{TelChu1,TongChu,XiChu,TelChu2,TelChu3}. In those works a generalized pseudospectral approach was used for the spatial discretization of the resonant Hamiltonian, and a non-Hermitian split-operator technique was implemented for the time-evolution operator. Telnov and Chu presented benchmark results for HHG  for monochromatic intense laser fields for the $\mbox{H}^{+}_{2}$ ion in \cite{TelChu2}. 

In the present article our goal is to report on HHG spectra for a linearly polarized intense laser field (with electric field aligned with the molecular axis) for the lowest two states of $\mbox{H}^{+}_{2}$, using the Floquet approach combined with a complex absorbing potential (CAP). This methodology was implemented before \cite{TsogMarko2} to calculate the ionization rate for the lowest two electronic states of the ion by strong continuous laser fields in the low-frequency limit. We compare our results for the HHG with those obtained in \cite{Zuo2} and \cite{TelChu2}. Our method differs from that of Ref.\cite{TelChu2} in that we do not use a time propagator, but assemble the wave function from the Floquet eigenstates.  

The organization of this paper is as follows. In Section II A we start with the basic theoretical methodology to solve the TDSE for the $\mbox{H}^{+}_{2}$ ion, while making use of the time periodicity. In Section II B we provide details of how to get HHG spectra within the non-Hermitian Floquet approach. Section III discusses our results for HHG spectra for the lowest two states of the $\mbox{H}^{+}_{2}$, which is followed by conclusions.      

\section{Theory }

\subsection{The Floquet Hamiltonian}

In the presence of an external field,  for a diatomic molecule, such as $\mbox{H}^{+}_{2}$ the induced electronic motion happens on a faster time scale than the nuclear motion. Thus, we treat the dynamics in the Born-Oppenheimer approximation, in which the two nuclei are fixed (on the au time scale), and only the electronic motion is taken into account.   
%The Coulomb two-center problem for one electron can be solved analytically \cite{Hyl,Jaffe,Bab}. 
The field-free electronic Hamiltonian of the $\mbox{H}^{+}_{2}$ molecule can be written in atomic units as
  \begin{equation}\label{H0}
 H_{0} = -\frac{1}{2} \nabla^{2}_{\mathbf{r}} - \frac{1}{|\mathbf{r}  + \frac{R}{2} \mathbf{e}_{z}|} - \frac{1}{|\mathbf{r}  - \frac{R}{2} \mathbf{e}_{z}|}, 
\end{equation}
where $\mathbf{r}$ is the electron position vector and $R$ is the internuclear separation. 

If we assume that the interaction of the electron with the external electric field $V_{L}(\mathbf{r},t)$ is periodic in time with period $T = 2\pi/\omega$, that is, $H(\mathbf{r},t + T) = H(\mathbf{r},t)$,  according to Floquet theory \cite{Floq, Sambe}, the solution $\Psi({\mathbf{r},t})$ to the time-dependent Schr\"odinger equation for the system 
\begin{eqnarray}\label{TDSE}%\nonumber
  i\frac{\partial}{\partial t}\Psi(\mathbf{r},t) = H(t)\Psi({\mathbf{r},t}) = [H_{0} + V_{L}(\mathbf{r},t)]\Psi({\mathbf{r},t}), 
\end{eqnarray}
can be written as:
%\begin{eqnarray}\label{TDSEsol}%\nonumber
%  \Psi({\mathbf{r},t}) = e^{(-i E_{F} t)} \Phi(\mathbf{r},t), \\
%  \Phi(\mathbf{r},t+T) = \Phi(\mathbf{r},t) = \sum^{\infty}_{n=-\infty} e^{i n \omega t} \phi_{n}(\mathbf{r}) \approx \sum^{N_{F}}_{-N_{F}} e^{i n\omega t} \phi_{n}(\mathbf{r}), 
% \Big[H(\mathbf{r},t) - i\frac{\partial}{\partial t} \Big] \Phi(\mathbf{r},t) = E_{F} \Phi(\mathbf{r},t),
%\end{eqnarray}
\begin{equation}\label{TDSEsol}%\nonumber
  \Psi({\mathbf{r},t}) = e^{(-i E_{F} t)} \Phi(\mathbf{r},t),
\end{equation}  
\begin{equation}\label{TDSEsol2}%\nonumber
  \Phi(\mathbf{r},t+T) = \Phi(\mathbf{r},t) = \sum^{\infty}_{n=-\infty} e^{i n \omega t} \phi_{n}(\mathbf{r}) \approx \sum^{N_{F}}_{-N_{F}} e^{i n\omega t} \phi_{n}(\mathbf{r}), 
% \Big[H(\mathbf{r},t) - i\frac{\partial}{\partial t} \Big] \Phi(\mathbf{r},t) = E_{F} \Phi(\mathbf{r},t),
\end{equation}  
where $E_{F}$ is called the Floquet quasi-energy, and the $ \phi_{n}(\mathbf{r})$ obey time-independent coupled-channel equations.  The last expression in (\ref{TDSEsol2}) represents the truncated ansatz used in practical calculations. 

Substitution of the solution ansatz (\ref{TDSEsol}) into the Schr\"odinger equation (\ref{TDSE}) leads to a time-dependent eigenvalue problem: 
\begin{eqnarray}\label{TDEVP}%\nonumber
 H_{F}(\mathbf{r},t) \Phi(\mathbf{r},t) = E_{F} \Phi(\mathbf{r},t), 
 \end{eqnarray}
where the Floquet Hamiltonian $H_{F}(\mathbf{r},t)$ is defined as
\begin{eqnarray}\label{HFl}%\nonumber
 H_{F}(\mathbf{r},t)  =  H(\mathbf{r},t) - i\frac{\partial}{\partial t}. 
 \end{eqnarray}
 
In this work we assume that the external field is provided by a linearly polarized monochromatic laser whose electric field is aligned with the internuclear axis of the $\mbox{H}^{+}_{2}$ ion, and that the dipole approximation is valid. Then the interaction $V_{L}(\mathbf{r},t)$ in length gauge takes the form
\begin{eqnarray}\label{intlg}%\nonumber
 V^{lg}_{L}(\mathbf{r},t) = F z \cos \omega t,  
 \end{eqnarray}
where $F$ is the laser field strength. 
%, and 
%\begin{eqnarray}\label{intvg}%\nonumber
% V^{vg}_{L}(\mathbf{r},t) = i \frac{F}{\omega} \sin \omega t \frac{\partial}{\partial z} + \frac{F^2}{2 \omega^2}\sin^2 \omega t 
% \end{eqnarray}
%in velocity gauge, where $F$ is the laser field strength. 
The length gauge is indeed more appropriate for low-frequency fields \cite{TsogMarko2}, and we employ this gauge for $\omega = 0.0428, 0.05695$, and $ 0.08565\,au$ in the present work, which correspond to wavelengths $\lambda = 1064$, $800$ and $532\,nm$, respectively. 

For the solution of the Floquet (\textit{steady-state}) Hamiltonian (\ref{HFl}), the time variable $t$ is treated in analogy to a coordinate variable, and the Schr\"odinger equation (\ref{TDEVP}) is solved as for the stationary states of the time-independent Schr\"odinger equation.  Once we find $\Phi(\mathbf{r},t)$ from the \textit{steady-state} Schr\"odinger equation (\ref{TDEVP}), we obtain the solution $\Psi(\mathbf{r},t)$ to the time-dependent Schr\"odinger equation (\ref{TDSE}) via equation (\ref{TDSEsol}).

We choose prolate spheroidal coordinates to deal with the $\mbox{H}^{+}_{2}$ ion (as described in \cite{TsogMarko2}), in which the Born-Oppenheimer treatment (equation (\ref{H0})) gives an analytic solution to the Schr\"odinger equation \cite{Hyl,Jaffe,Bab}.

The final time-independent coupled equations for the non-Hermitian matrix problem in length gauge to be implemented are
\begin{eqnarray}\label{Flomateqlg}\nonumber
 [ H_{0}(\mu,\nu) - i\eta W(\mu)] \phi_{n}(\mu,\nu) + \frac{1}{2} F z [\phi_{n-1} + \phi_{n+1}] = (E_{F} - n\omega) \phi_{n}(\mu,\nu).\\
 (n = 0, \pm 1, \pm 2,\ldots, \pm N_{F})
 \end{eqnarray}
 Details of the discretization of this equation and calculation of the resonance parameter $E_{F}$ can be found in \cite{TsogMarko2}. 

The field-free Hamiltonian (\ref{H0}) and the complex absorbing potential $W(\mu) $ in equation (\ref{Flomateqlg}) are given as
\begin{eqnarray}\label{Hpsphc}\nonumber
  H_{0} =  -\frac{1}{2} \frac{4}{R^2 (\mu^2 - \nu^2)} \Big[ \frac{\partial}{\partial \mu}[(\mu^2-1) \frac{\partial}{\partial \mu}]  + \frac{\partial}{\partial \nu}[(1-\nu^2) \frac{\partial}{\partial \nu}]  + \\\frac{\mu^2 - \nu^2}{(\mu^2-1)(1-\nu^2)} \frac{\partial^2}{\partial \varphi^2} \Big] - \frac{4\mu}{R (\mu^2 - \nu^2)},  
 \end{eqnarray}
and
\begin{equation}\label{ACStHam2}
% H  =  H_{res}(\mu,\nu,t) - i \eta W(\mu), \quad W(\mu) = \Theta (\mu-\mu_{c}) (\mu - \mu_{c})^{2}, 
 W(\mu) = \Theta (\mu-\mu_{c}) (\mu - \mu_{c})^{2}, 
\end{equation}
where $\Theta$ is the Heaviside step function, $\eta$ is a small positive parameter, and $\mu_{c}$ determines the ellipse outside of which the CAP dampens the outgoing wave in the asymptotic  region.

\subsection{Calculation of HHG spectra from the non-Hermitian Floquet approach}

Once we have found the time-dependent Floquet wave function $\Psi(\mathbf{r},t)$ via equation (\ref{TDSEsol}), we can compute the time-dependent dipole moment along the internuclear axis, $d(t)$, as
\begin{eqnarray}\label{dipmom}%\nonumber
 d(t) = \langle \Psi (\mathbf{r},t) | z | \Psi (\mathbf{r},t) \rangle\,.
\end{eqnarray}
Following \cite{TelChu2, Land}, the $n$th-order harmonic generation rates $\Gamma_{n}$ (the number of photons with frequency $n\omega$ emitted per unit time) are calculated by the Larmor formula 
\begin{eqnarray}\label{partrate}%\nonumber
 \mathbf{\Gamma}_{n} = \frac{4 n^{3} \omega^{3}}{3 c^{3}} |d_{n}|^2, 
\end{eqnarray}
where $c$ is the speed of light, and $d_{n}$ is the Fourier transform of the time-dependent dipole moment (\ref{dipmom}) as
\begin{eqnarray}\label{ftrnsdipmom}%\nonumber
  d_{n} = \frac{1}{T}\int^{T}_{0} dt \exp(i n \omega t)\, d(t). 
\end{eqnarray}
As shown by Telnov and Chu \cite{TelChu2} in Floquet theory the HHG rates are obtained to the same accuracy irrespective of whether one uses the dipole operator (\ref{dipmom}) or the velocity or acceleration forms. 

\section{Results and Discussion}

\subsection{ HHG rates for the equilibrium separation $R=2\,au$}

We attempted first to obtain the HHG spectra previously reported by Telnov and Chu \cite{TelChu2}. In our grid representation we have two main parameters, and the results for the harmonic generation spectra appear to be sensitive to them, implicitly via the Floquet wave function $\Psi(\mu,\nu,t)$. One of two (artificial) parameters which control the wave function is $\mu_{c}$, which determines the region where the CAP starts. Another one is the absorbing strength parameter $\eta$. Ideally, the results ought to be insensitive to these two parameters. It is obvious that the value of $\mu_{c}$ should be larger than the quiver radius of a free electron $\alpha_{0} = F/\omega^2$ (in atomic units), because the main contribution to the harmonic generation spectra comes from the free electron driven back to its parent ion or two-center core.   

In analogy to the $\eta$-trajectory in the calculation of the resonance parameter $E^{(0)}_{F}$ in \cite{TsogMarko1}, we initially obtain the HHG rate $\Gamma_{n}$ for varying $\eta^{(0)}$. Within a certain range of the $\eta^{(0)}$-trajectory, namely where the complex eigenenergy value $E^{(0)}_{F}$ stabilizes, the resonance wave function is accurate, and in turn, it should yield accurate HHG spectra there. 

\begin{figure}[h]
\centering
%  \includegraphics[width=0.55\textwidth]{Ad_diab_en_F0p0533_lambda288.eps}
%  \includegraphics[width=0.50\textwidth]{h2p_ion_rate_1su.pdf}
    \mbox{\subfigure[]{\includegraphics[width=0.330\textwidth]{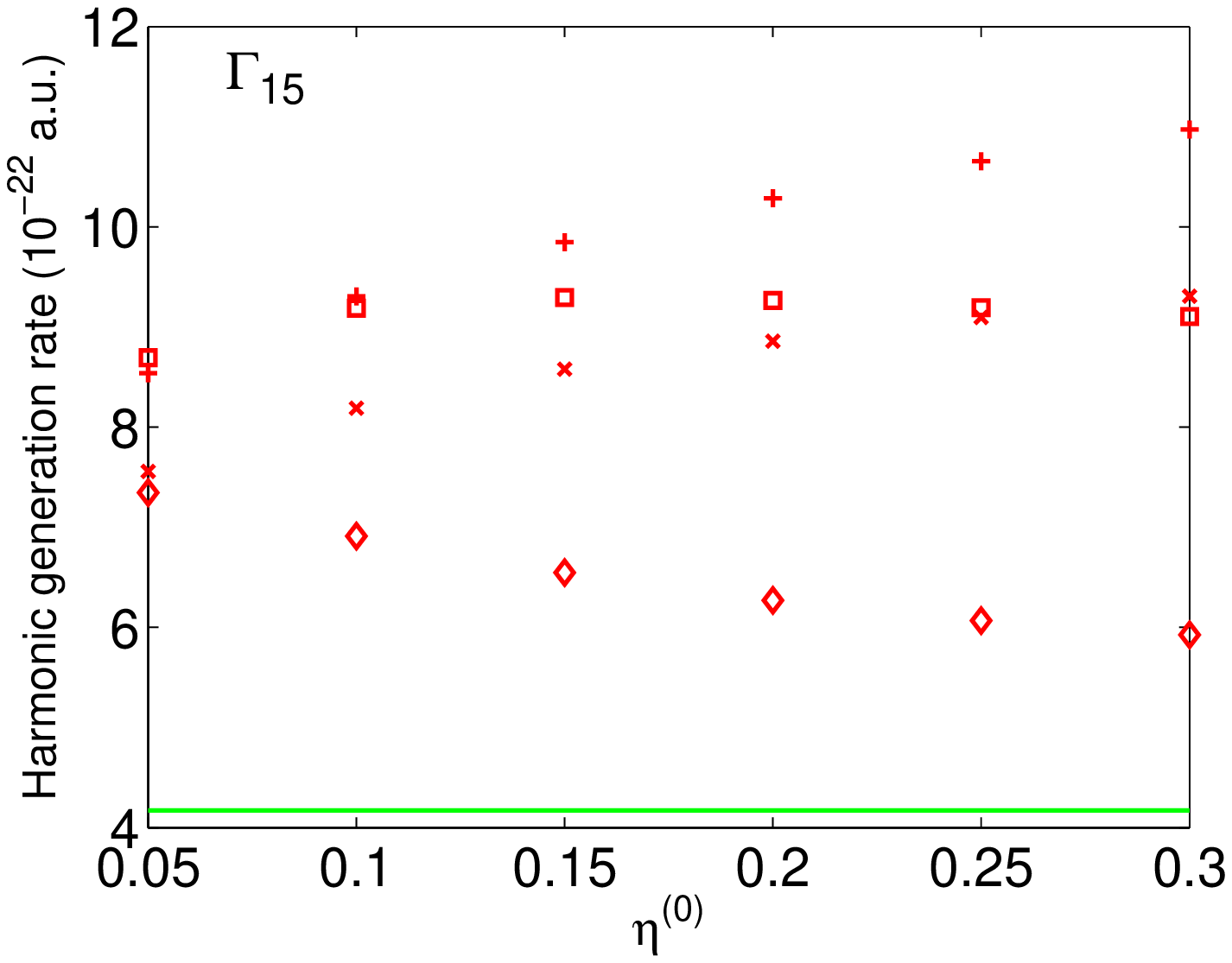}}
      \subfigure[]{\includegraphics[width=0.330\textwidth]{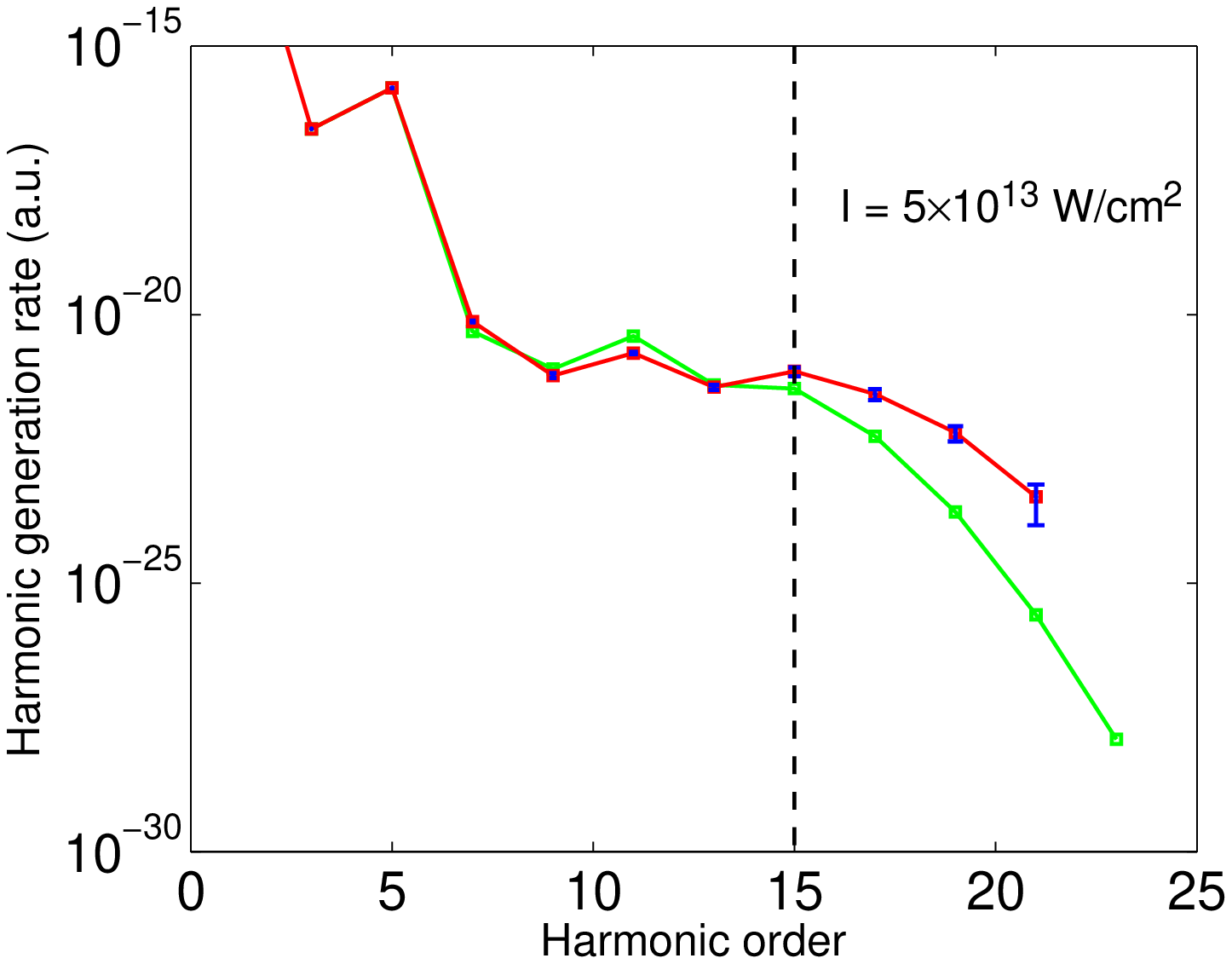}}
      \subfigure[]{\includegraphics[width=0.330\textwidth]{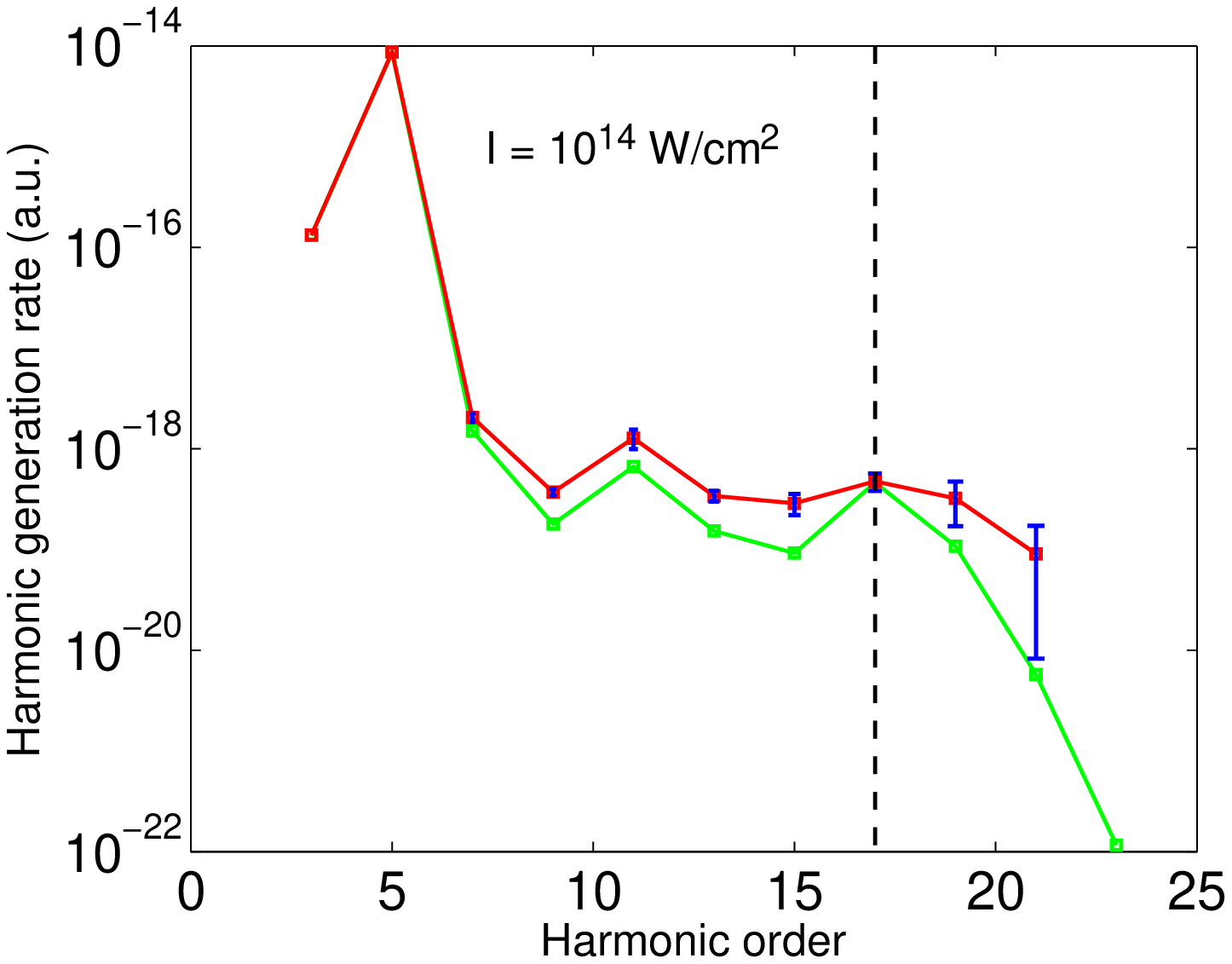}} 
%      \subfigure[]{\includegraphics[width=0.280\textwidth]{Pade6_HHG_n31_R2_low_F0p1194_w0p0856_eta0p168_muc_val.eps}} 
      }

\caption{ The HHG rate for the lower state of $\mbox{H}^{+}_{2}$ at internuclear separation $R=2\,au$, and $\lambda = 532\,nm$: (a) $I=5\times 10^{13}\,W/cm^{2}$, $\Gamma_{15}$ rate vs $\eta^{(0)}$ for various values of $\mu_{c}$, namely 13.5 (diamonds), $14.5$ (squares), $15.5$ (plus signs) and $16.5\,a.u.$ (crosses); the green line shows the value from Telnov and Chu \cite{TelChu2}; (b) plot for more harmonics than shown in (a); hereΠ$N_{max} \approx 15$; (c) $I=1\times 10^{14}\,W/cm^{2}$, and $N_{max}\approx 17$. The green line in (b, c) connects the data of \cite{TelChu2} while the red line connects the present data. The blue error bars are based on the calculations with different $\mu_{c}$. The vertical dashed lines indicate the semi-calssical cutoff values $N_{max}$. }  % -0.506673584605708 - 0.009964962098731*I:
\label{fig:1}       % Give a unique label
\end{figure}

In panel (a) of Fig.~1 we show the HHG rate $\Gamma_{15}$ vs $\eta^{(0)}$ for the lower state of the $\mbox{H}^{+}_{2}$ ion at the equilibrium separation $(R=2\,au)$ in the field of intensity $I=5\times 10^{13}\,W/cm^{2}$ and wavelength $\lambda = 532\,nm$. In panel (a) we show results for $\Gamma_{15}$ for $\eta^{(0)} \geq 0.05$ (at $\eta^{(0)} \leq 0.05$ the computation is inaccurate). We chose four different $\mu_{c}$ values, in the range $13.5\leq \mu_{c} \leq 16.5\,au$, which are much larger than $\alpha_{0} = 5.15\,au$. The bottom green line shows the result obtained by Telnov and Chu \cite{TelChu2}. Our results for $\Gamma_{15}$ are higher by up to a factor of $2.5$, as compared to with the value $\Gamma_{15} = 4.17\times10^{-22}\,au$ given in \cite{TelChu2}.  In panel (b) we show the HHG spectrum. We note that for orders $n<15$ the agreement with the results of \cite{TelChu2} is good and is independent of the chosen value of $\mu_{c}$. In all plots of the HHG rate we do not show results for order $n=1$, because they are usually much higher. The cutoff position is around $n=15$, which is indeed consistent with our result. For each harmonic order we use four different values of $\mu_{c}$ to compute $\Gamma_{n}$, and use them to define an average value with standard deviation. Panel (c) shows the same plot for the HHG rate for the doubled intensity, $1\times 10^{14}\,W/cm^{2}$. The cutoff law is clearly obeyed around $n = 17$ by both the present and previous \cite{TelChu2} results.  As compared to Ref.\cite{TelChu2} our HHG rates are higher above the cutoff but also less certain.  

\begin{figure}[h]
\centering
%  \includegraphics[width=0.55\textwidth]{Ad_diab_en_F0p0533_lambda288.eps}
%  \includegraphics[width=0.50\textwidth]{h2p_ion_rate_1su.pdf}
    \mbox{\subfigure[]{\includegraphics[width=0.330\textwidth]{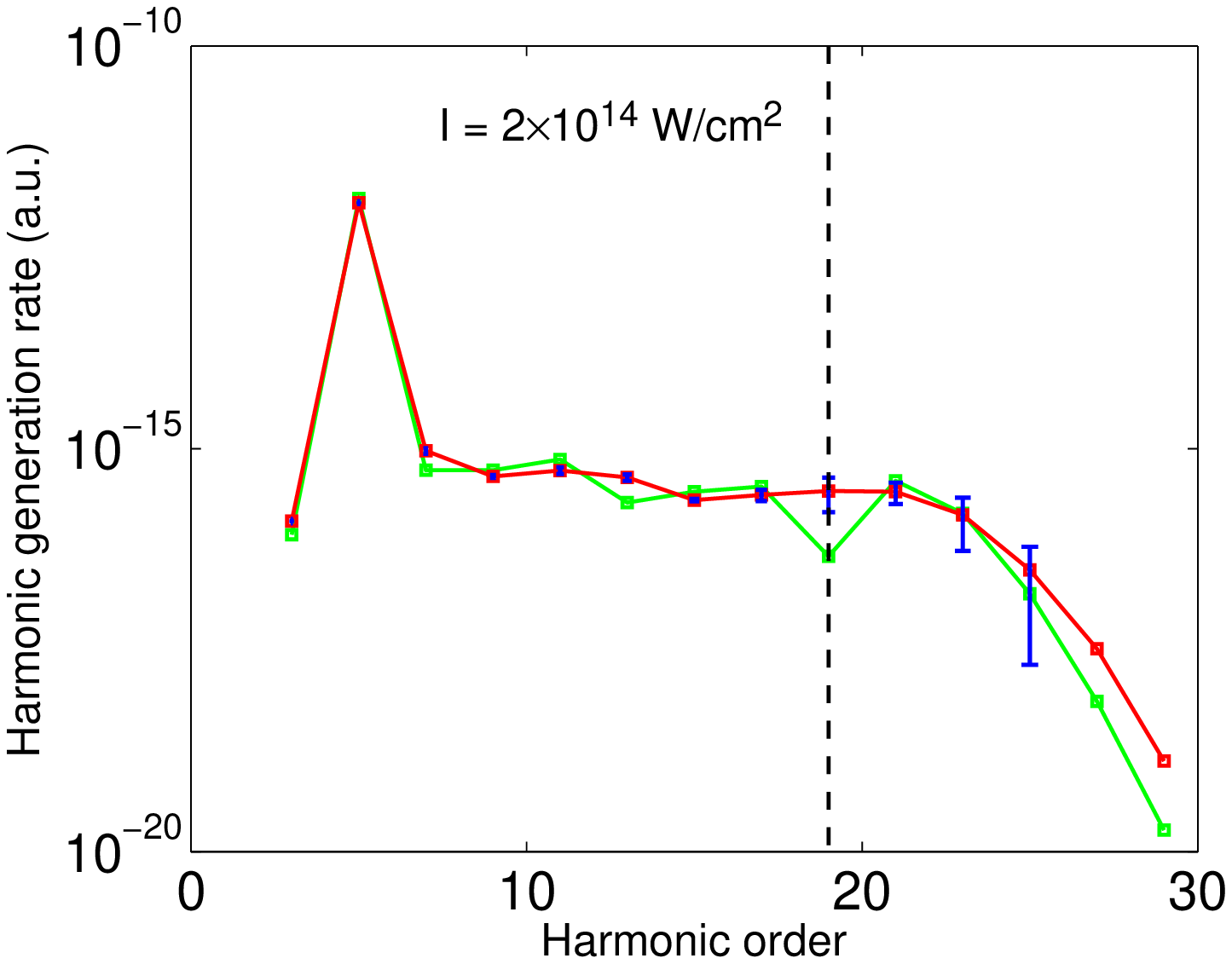}}
      \subfigure[]{\includegraphics[width=0.330\textwidth]{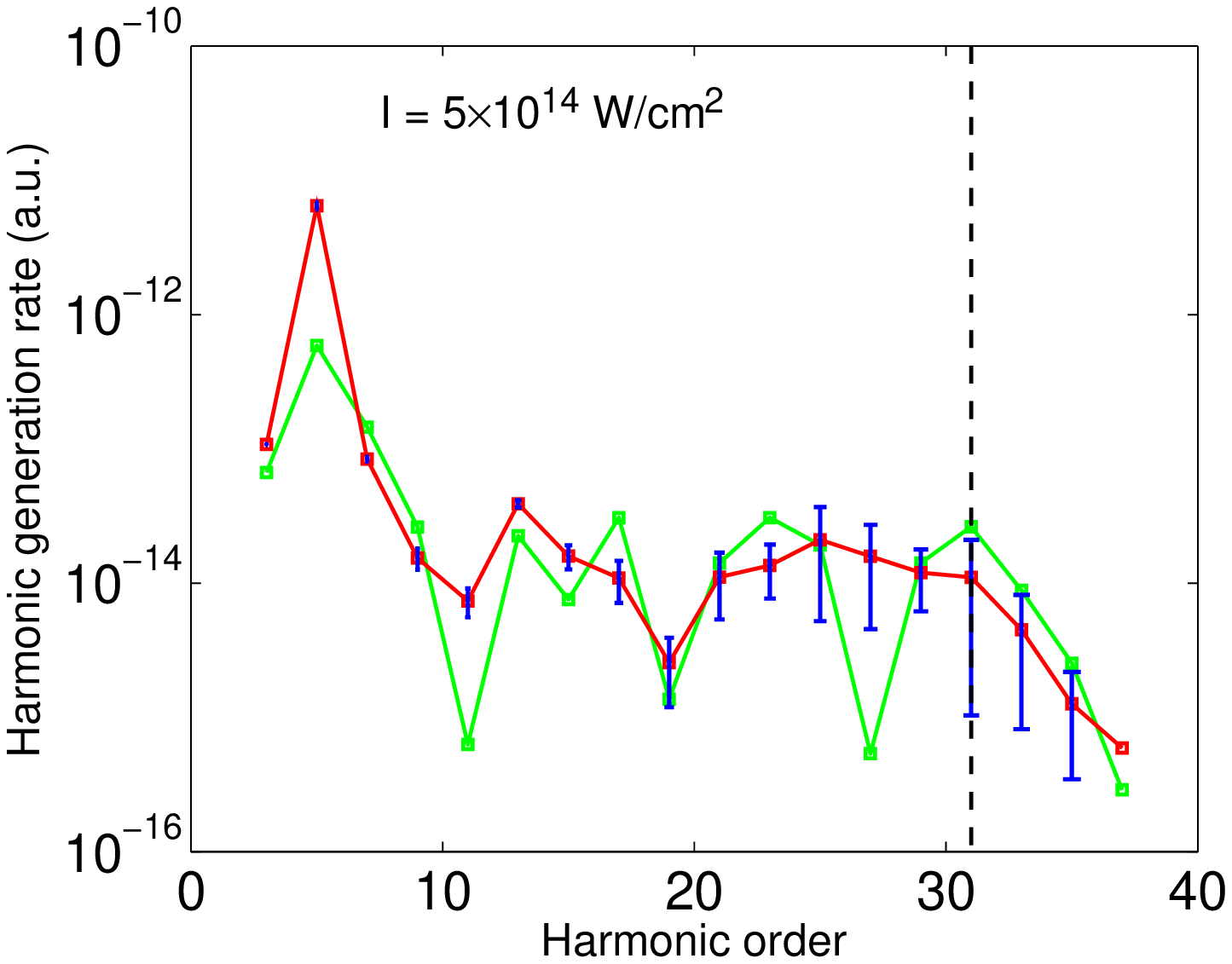}}
      \subfigure[]{\includegraphics[width=0.330\textwidth]{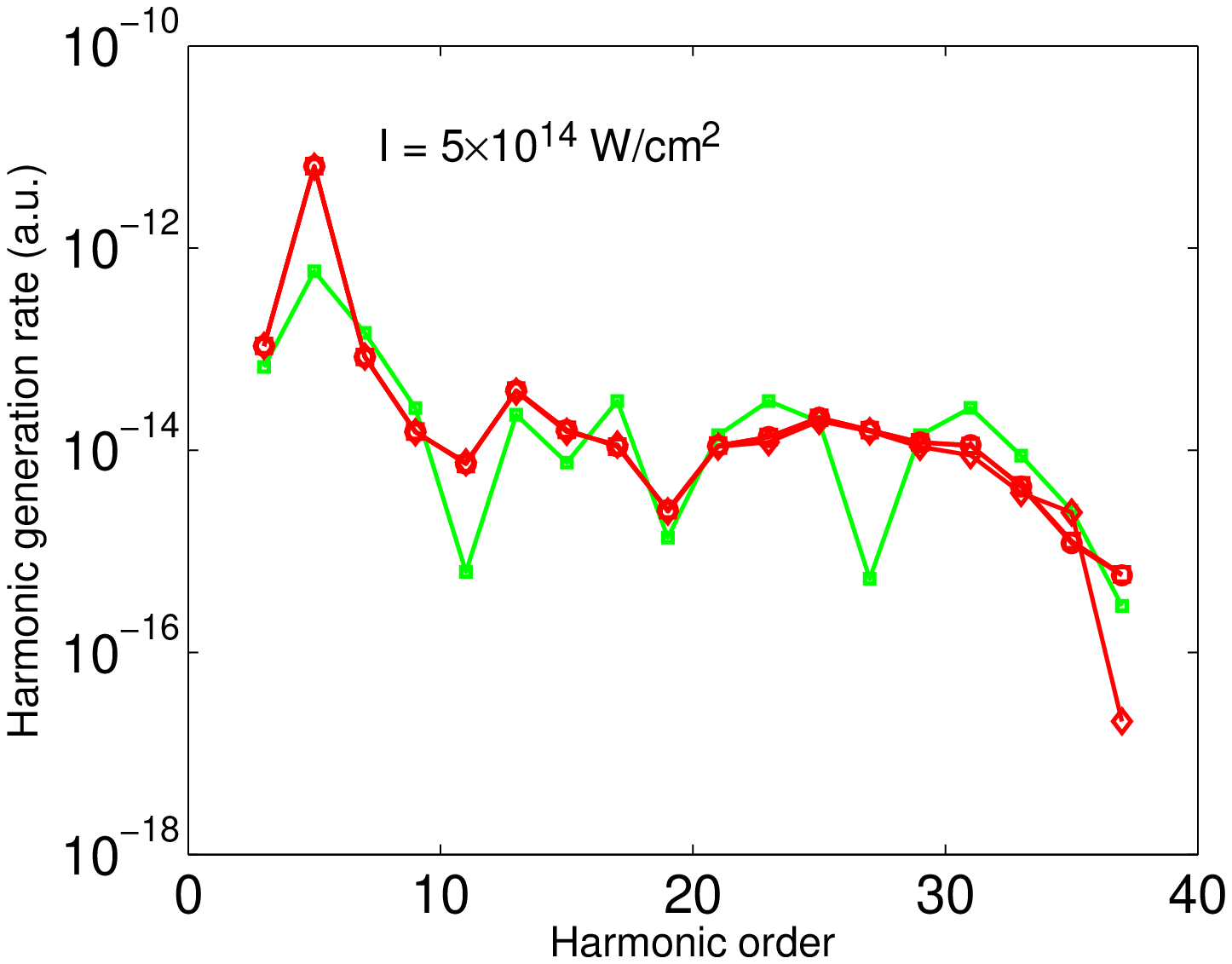}} 
%      \subfigure[]{\includegraphics[width=0.280\textwidth]{Pade6_HHG_n31_R2_low_F0p1194_w0p0856_eta0p168_muc_val.eps}} 
      }

\caption{The same plot as shown in Figure 1\,(b, c), however the laser intensities are $I=2\times 10^{14}\,W/cm^{2}$ (a), and $I=5\times 10^{14}\,W/cm^{2}$ (b). In (c) the same plot is shown as in (b), but the number of Floquet channels is chosen to be $N_{F} = 72$ (diamonds), $82$ (squares) and $86$ (circles), respectively. The classical cutoff positions are around $19$ (a), and $31$ (b, c) and are indicated by vertical dashed lines.}  % -0.506673584605708 - 0.009964962098731*I:
\label{fig:1}       % Give a unique label
\end{figure}

In Fig.~2 data are shown for increased laser intensities. In panel (a) of Fig.~2 in the HHG spectra the cutoff position moves up to $19$, but the calculated spectrum extends the plateau to higher orders.  Our result agrees well with that obtained by Telnov and Chu \cite{TelChu2}. In panel (b) of Fig.~2 we show the same plot for $I=5\times 10^{14}\,W/cm^{2}$. In this case although the general features of the obtained HHG spectrum follow those of Ref.\cite{TelChu2}, it does show significant deviation at certain harmonic orders, namely for $n=5, 11$ and $27$. To check our answer carefully, we gradually increased the number of Floquet channels, because the HHG rates at higher harmonic orders require a higher number of photon couplings. In panel (c) of Fig.~2 we display the same plot as shown in panel (b), but the number of Floquet channels $N_{F}$ is $72$, $82$ and $86$, respectively. Thus we know that the results are converged at $N_{F}=86$ in the HHG order range presented in our plots. The data Fig.~2c  are based on a matrix diagonalization with $N_{F}=86$, with the specified truncation imposed in eqn (\ref{TDSEsol}) when computing (\ref{dipmom}).         

\begin{figure}[h]
\centering
%  \includegraphics[width=0.55\textwidth]{Ad_diab_en_F0p0533_lambda288.eps}
%  \includegraphics[width=0.50\textwidth]{h2p_ion_rate_1su.pdf}
    \mbox{\subfigure[]{\includegraphics[width=0.330\textwidth]{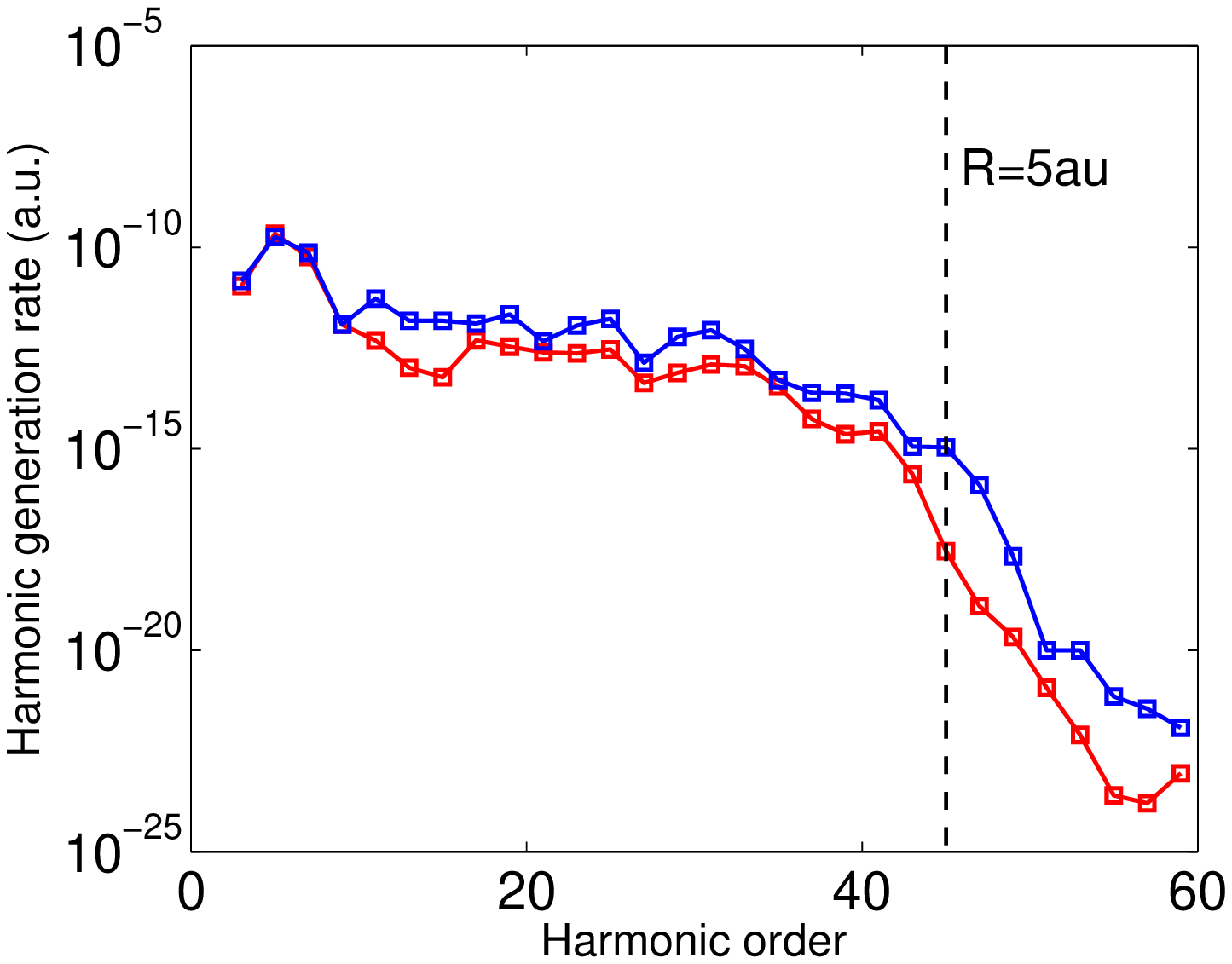}}
      \subfigure[]{\includegraphics[width=0.330\textwidth]{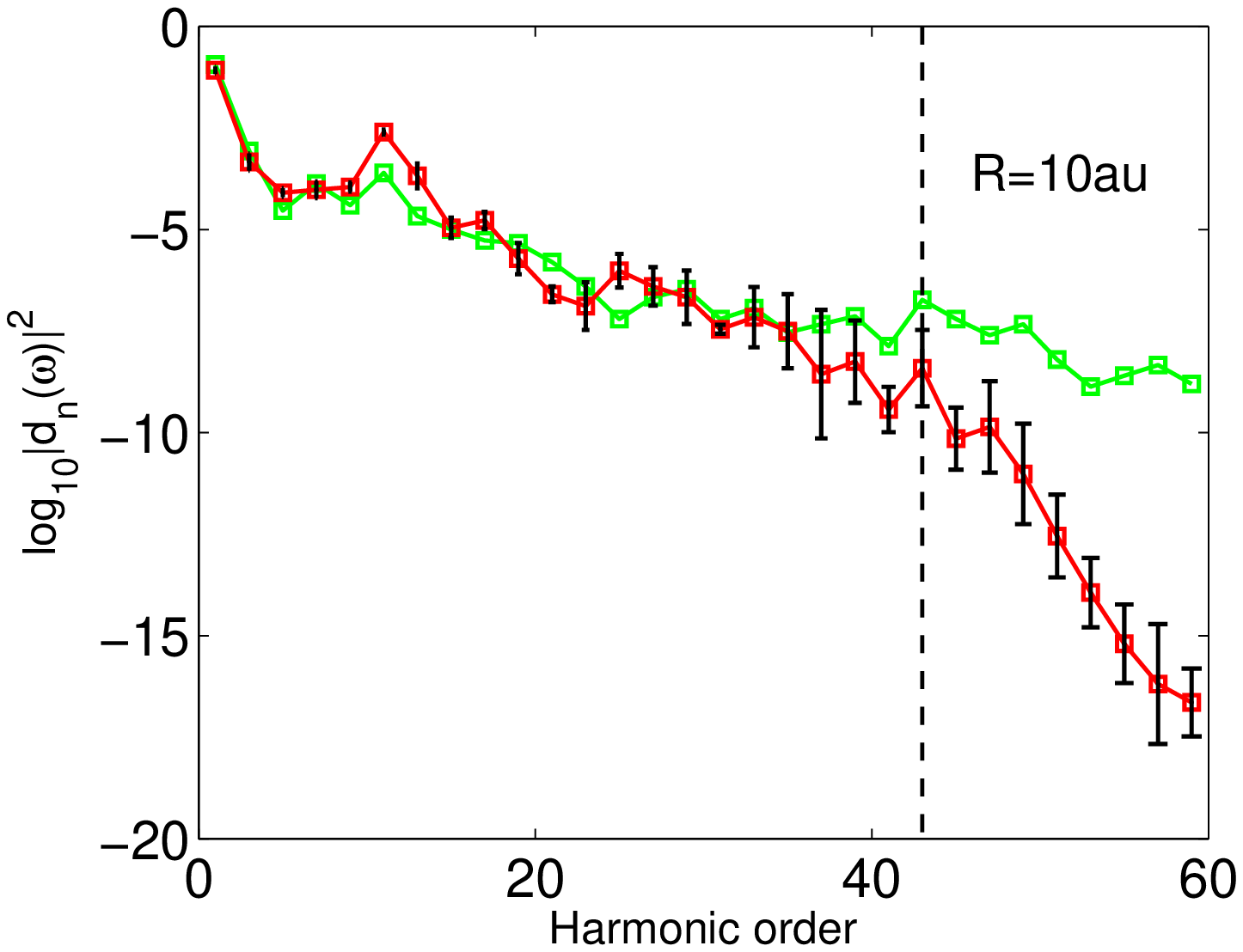}}
      \subfigure[]{\includegraphics[width=0.330\textwidth]{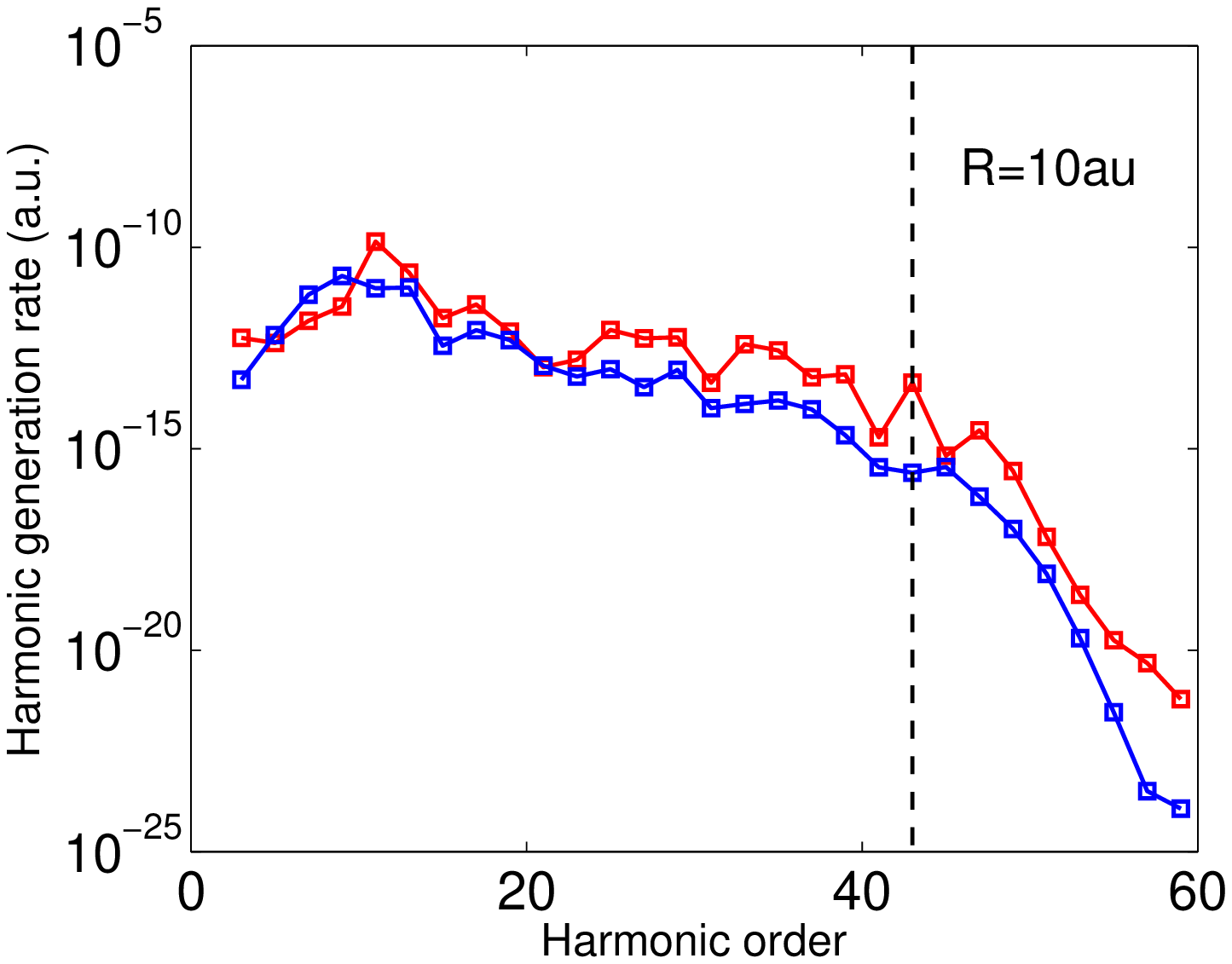}} 
%      \subfigure[]{\includegraphics[width=0.280\textwidth]{Pade6_HHG_n31_R2_low_F0p1194_w0p0856_eta0p168_muc_val.eps}} 
      }

\caption{HHG rates for the lower (red) and and upper (blue) states of the $\mbox{H}^{+}_{2}$ ion at $R = 5\,au$ (a) and $R = 10\,au$ (c). Panel (b) shows the HHG spectrum for the lower state (red) at $R=10\,au$ to be compared with the result obtained in \cite{Zuo2} (green). The laser field parameters are $I=1\times 10^{14}\,W/cm^{2}$ and $\lambda = 1064\,nm$.  }  % -0.506673584605708 - 0.009964962098731*I:
\label{fig:1}       % Give a unique label
\end{figure}

Next we continue with HHG rates while moving towards the low-frecuency limit. In Fig.~3 we show the rates for the lower (red) and upper (blue) states for the $\mbox{H}^{+}_{2}$ ion for laser fields of $I=1\times 10^{14}\,W/cm^{2}$ and $\lambda = 1064\,nm$. All HHG calculations are carried out with $\eta^{(0)} = 0.25$. Panel (a) shows the spectrum at internuclear separation $R=5\,au$, while panels (b, c) demonstrate corresponding results at $R=10\,au$. Given that $\omega = 0.0428\,au$, the cutoff positions given by the classical formula (\ref{cutoffn}) are found around $n=45$ and $n=43$ at $R=5\,au$ and $R=10\,au$, respectively, and are shown by vertical dashed lines for the lower state in Figure 3 (for the upper state, the cutoff position is close to it, since both states have almost the same ionization potential at large internuclear separations). As  discussed by Bandrauk and co-workers in \cite{Zuo1,Zuo2}, these classical cutoff positions are referred to as the atomic plateau. They argue that a first plateau region can be identified as a molecular plateau: its cutoff occurs at harmonic order 
\begin{eqnarray}\label{nMol}%\nonumber
%  n_{M}= 2 \Omega_{R}/\omega = 2 d_{0} F/\omega \approx \frac{R F}{\omega} ,
  n_{M}= \frac{2 \Omega_{R}}{\omega} = \frac{2 d_{0} F}{\omega} \approx \frac{R F}{\omega} ,  
\end{eqnarray}
where $\Omega_{R}$ is the Rabi frequency for driving transitions between the $1\sigma_{g}$ and $1\sigma_{u}$ states and the transition dipole moment $d_{0}$ grows towards $R/2$ with increasing $R$ (for details cf \cite{Zuo2}). According to this model in which the two lowest states are driven resonantly, since $\omega\geq \epsilon_{1\sigma_{u}} - \epsilon_{1\sigma_{g}}$, the values of the cutoff positions can be found at $n_{M}=5$ for $R = 5\,au$, and $n_{M}=11$ for $R=10\,au$, respectively. These calculated values of $n_{M}$ can be observed in the data given in Figure 3.  Panel (b) of Figure 3 shows the HHG spectrum calculated by equation (\ref{ftrnsdipmom}) for the lower state of the ion and its comparison with that obtained in \cite{Zuo2} (green). We note that the agreement between the Floquet result and the calculation for a finite $30$- cycle pulse is excellent up to order $39$. Beyond this order the harmonics for the finite pulse with square envelope continue to be strong, while the Floquet results fall off. 

\begin{figure}[h]
\centering
%  \includegraphics[width=0.55\textwidth]{Ad_diab_en_F0p0533_lambda288.eps}
%  \includegraphics[width=0.50\textwidth]{h2p_ion_rate_1su.pdf}
    \mbox{\subfigure[]{\includegraphics[width=0.310\textwidth]{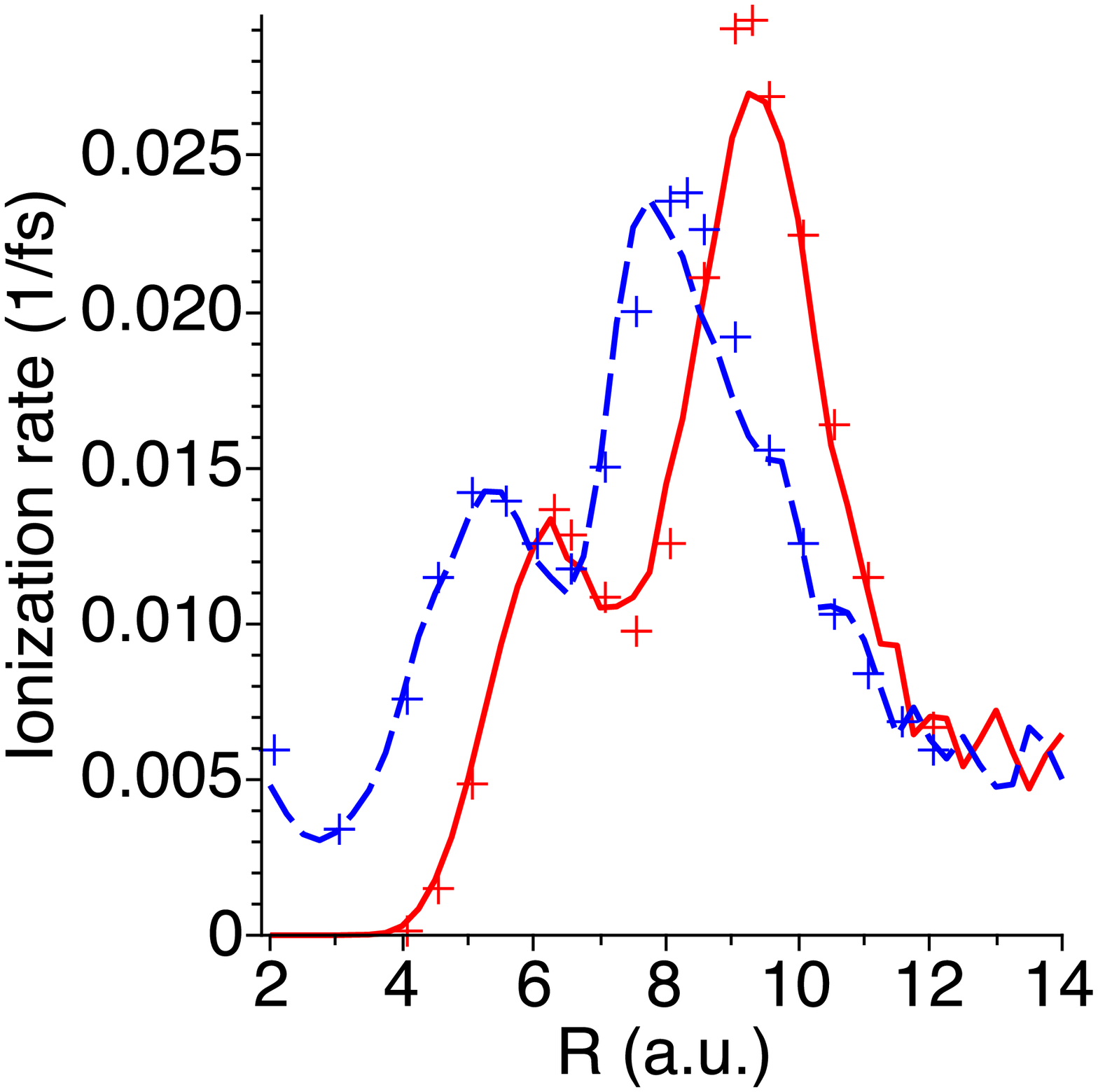}}
      \subfigure[]{\includegraphics[width=0.310\textwidth]{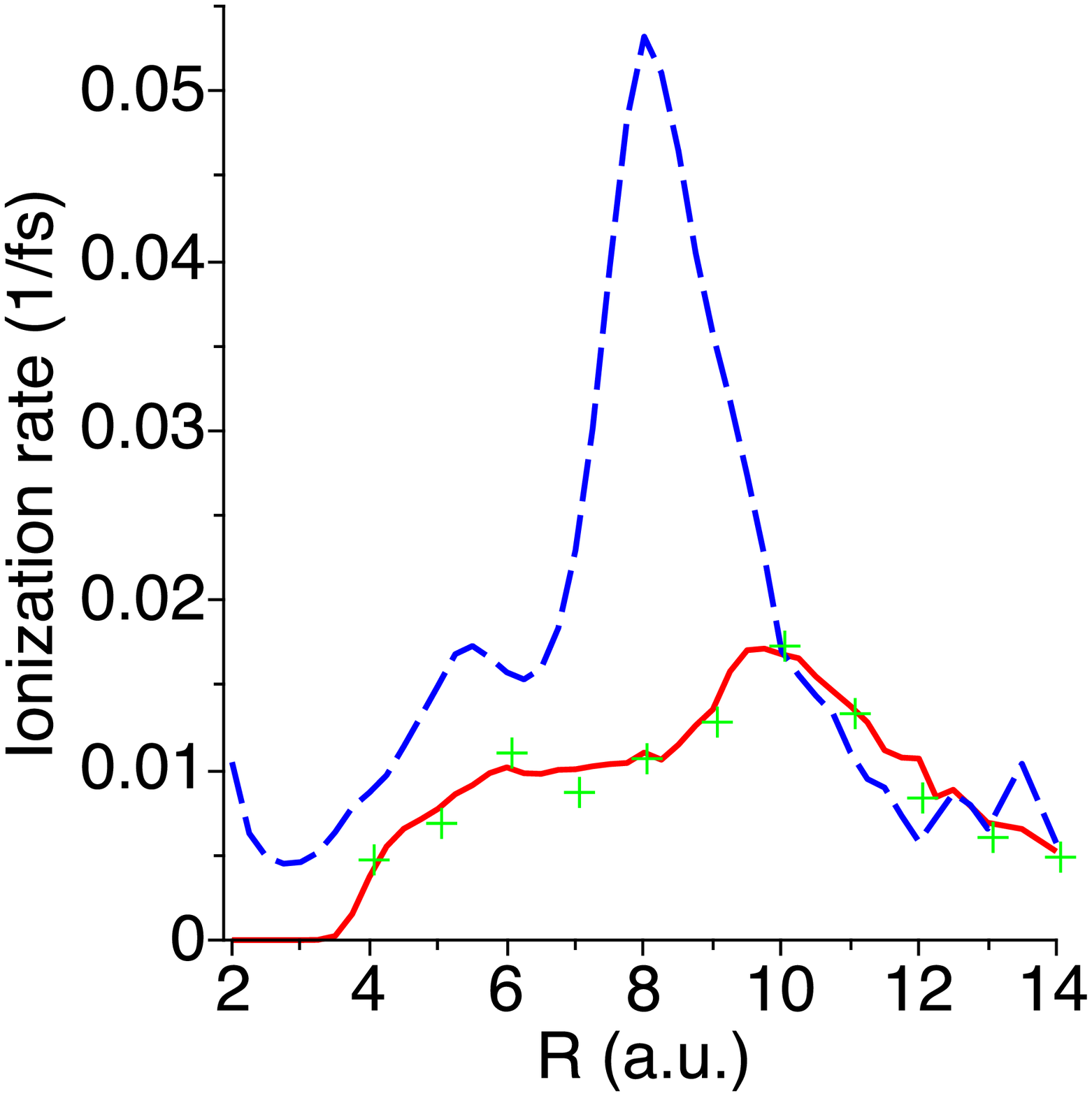}}
%      \subfigure[]{\includegraphics[width=0.350\textwidth]{HHG_F0p0533_w0p0428_eta0p251_Nt256_Gn11.eps}} 
%      \subfigure[]{\includegraphics[width=0.28\textwidth]{HHG_F0p0533_w0p0428_eta0p251_Nt256_Gn15.eps}} 
      }

\caption{Ionization rate (in fs$^{-1}$) as a function of $R$ for the lower and upper states of $\mbox{H}^{+}_{2}$. Curves: present results, solid red for the lower state, and dashed blue for  the upper state;  crosses: Chu et al \cite{Xichu2} (red) (a), and Bandrauk and Lu \cite{Zuo2} (green) (b). The field parameters are $I=1\times 10^{14}\,W/cm^{2}$ and $\lambda = 1064\,nm$ (a) and  $\lambda = 800\,nm$ (b) respectively.}  % -0.506673584605708 - 0.009964962098731*I:
\label{fig:1}       % Give a unique label
\end{figure}

\subsection{Harmonic generation rates as a function of internuclear separation $R$}

In this subsection we present HG rates for moderate orders $n$, i.e., in the molecular plateau region at the intensity $1\times 10^{14}\,W/cm^{2}$ as a function of internuclear separation $R$. In Figure 4 we show the ionization rates for the lower and upper states of the ion for two wavelengths of the laser field, $1064\,nm$ and $800\,nm$. Since the physical interpretation for the enhancement of the ionization rates shown in Figure 4 is discussed in \cite{Zuob,Plum,Mul,Mad} and later in \cite{TsogMarko1,TsogMarko2}, we do not repeat it here. Our goal is to demonstrate how the enhancement of the ionization rate in certain $R$-regions affects the HG rates within the region of the molecular plateau, i.e., for $n\leq n_{M}$ (\ref{nMol}).

\begin{figure}[h]
\centering
%  \includegraphics[width=0.55\textwidth]{Ad_diab_en_F0p0533_lambda288.eps}
%  \includegraphics[width=0.50\textwidth]{h2p_ion_rate_1su.pdf}
      \subfigure[]{\includegraphics[width=0.320\textwidth]{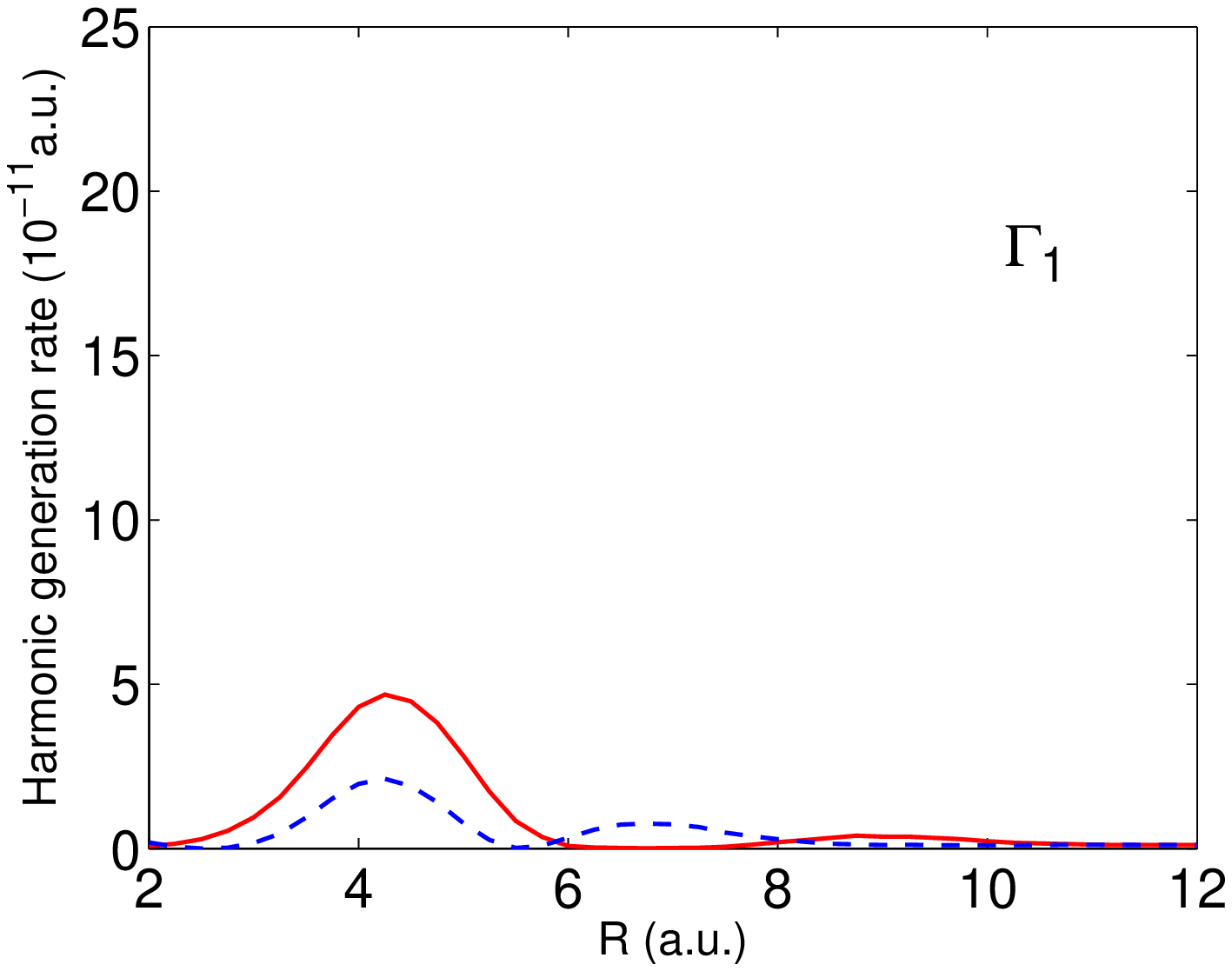}}
      \subfigure[]{\includegraphics[width=0.320\textwidth]{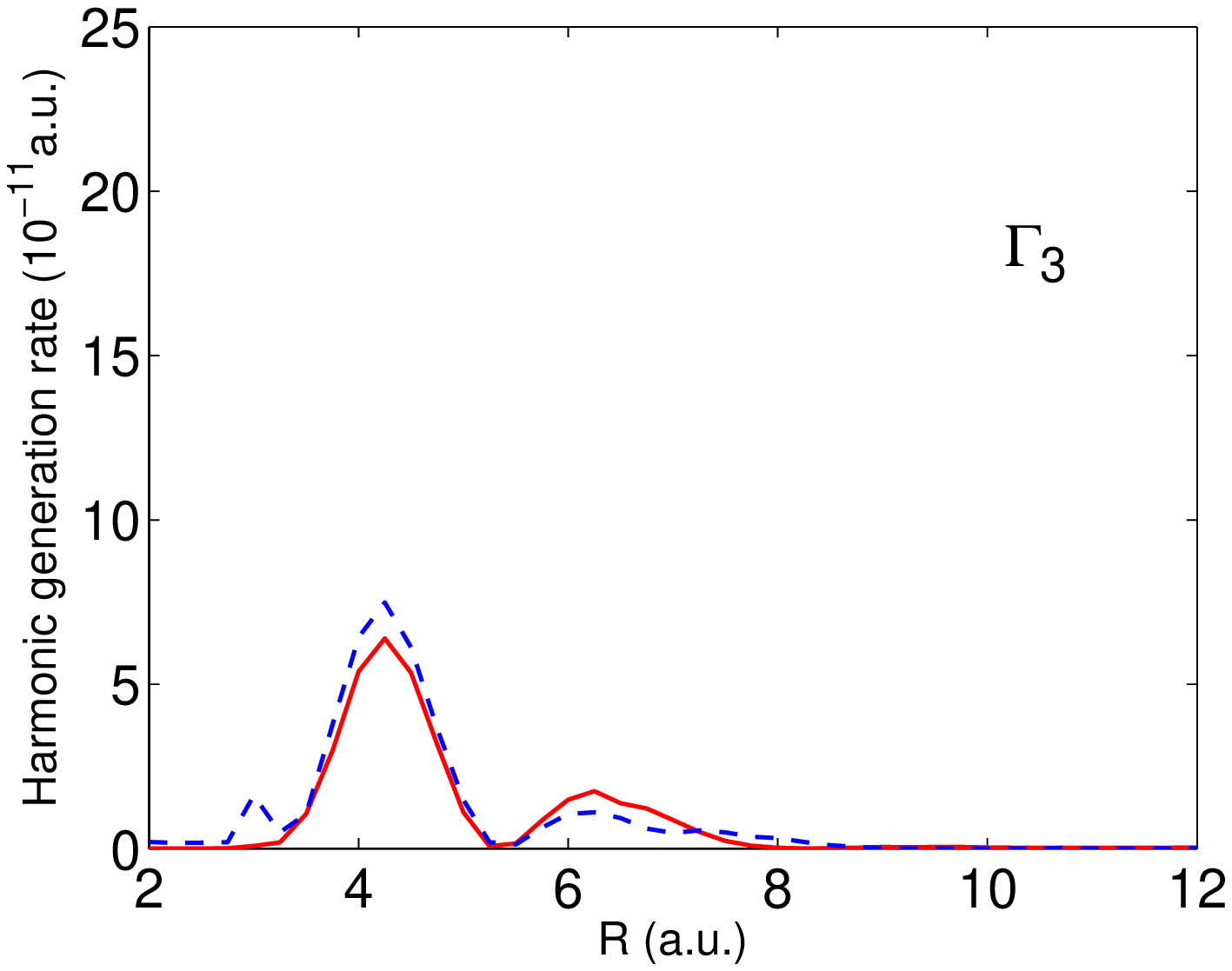}}
      \subfigure[]{\includegraphics[width=0.320\textwidth]{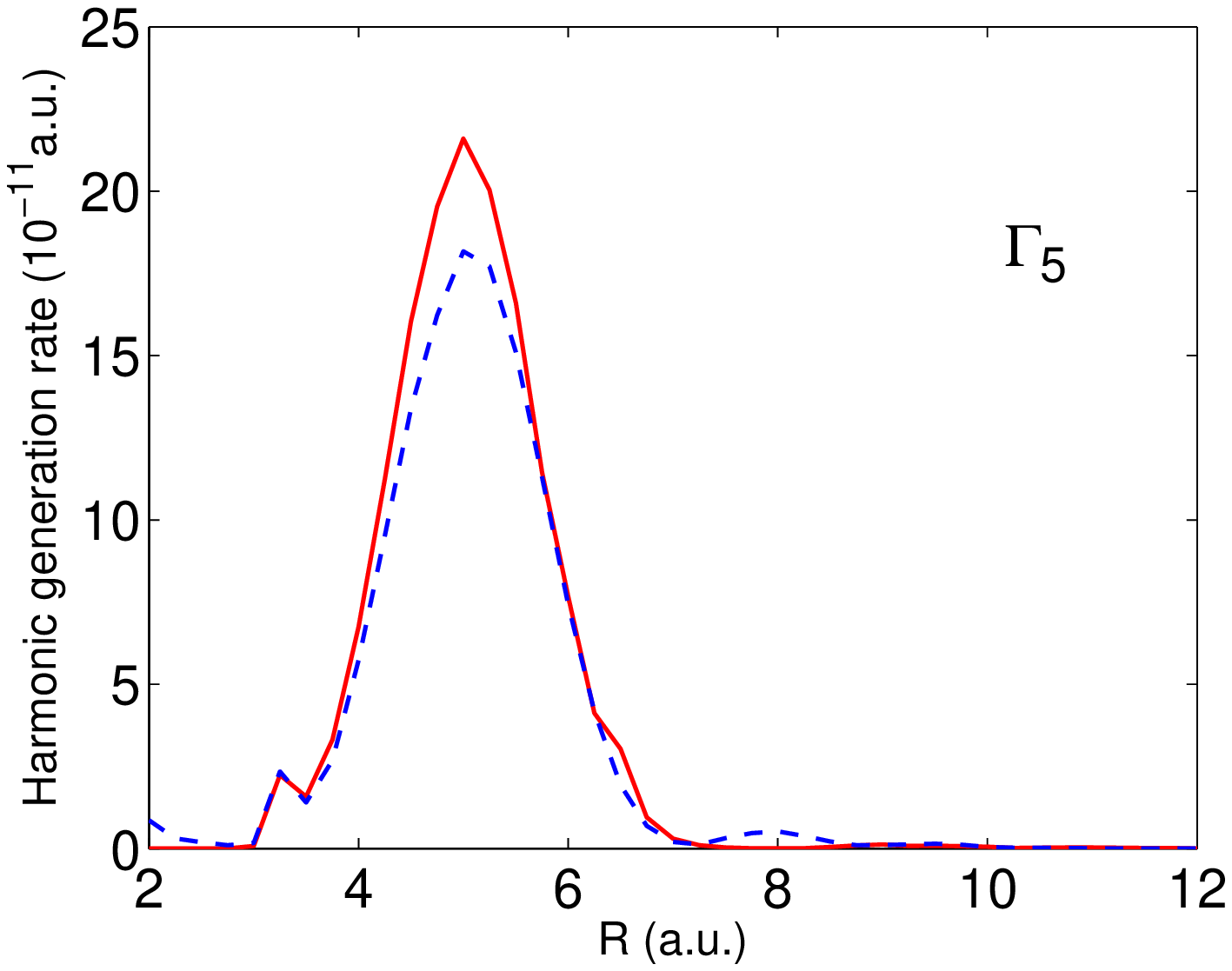}}
      \subfigure[]{\includegraphics[width=0.320\textwidth]{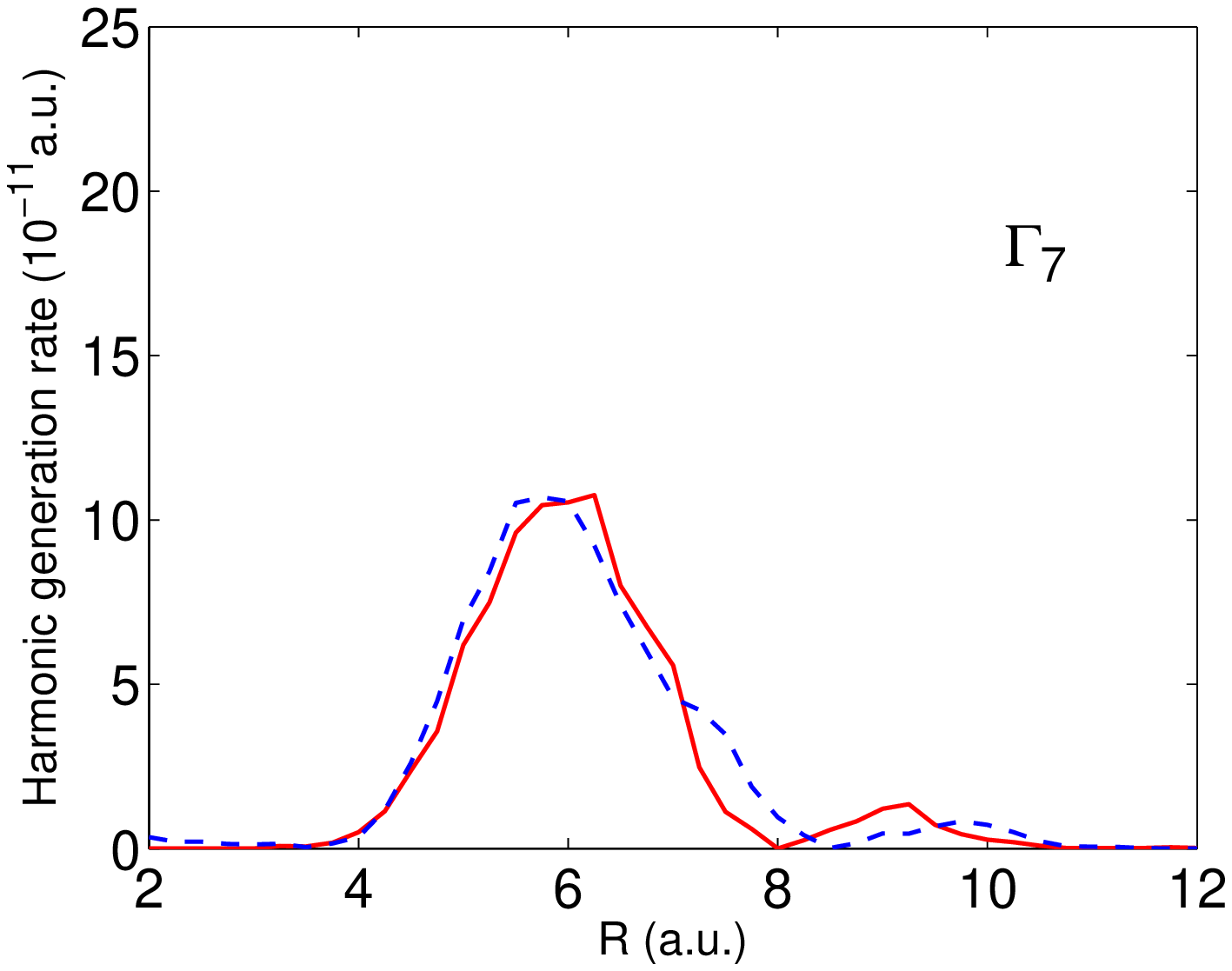}}
      \subfigure[]{\includegraphics[width=0.320\textwidth]{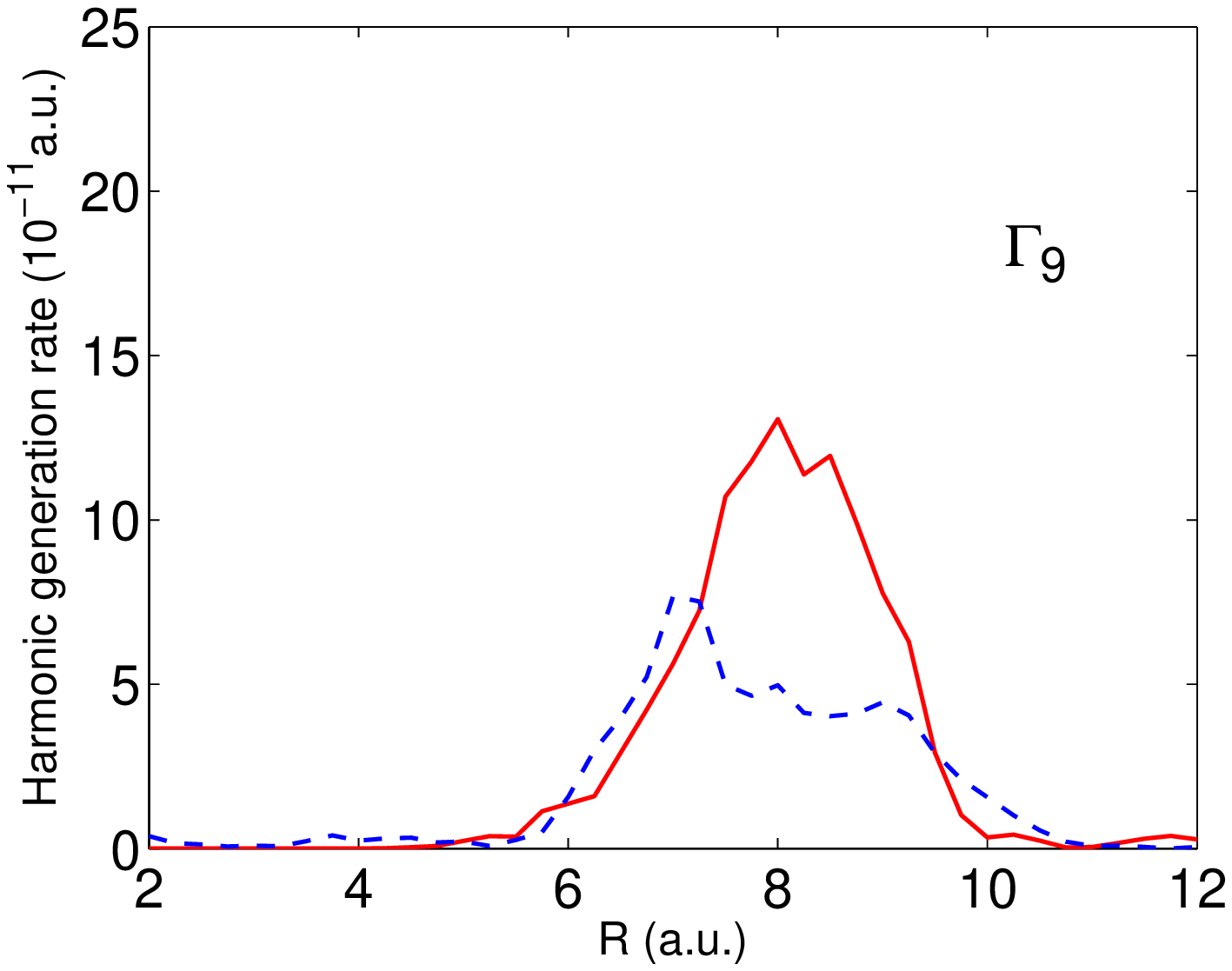}}
      \subfigure[]{\includegraphics[width=0.320\textwidth]{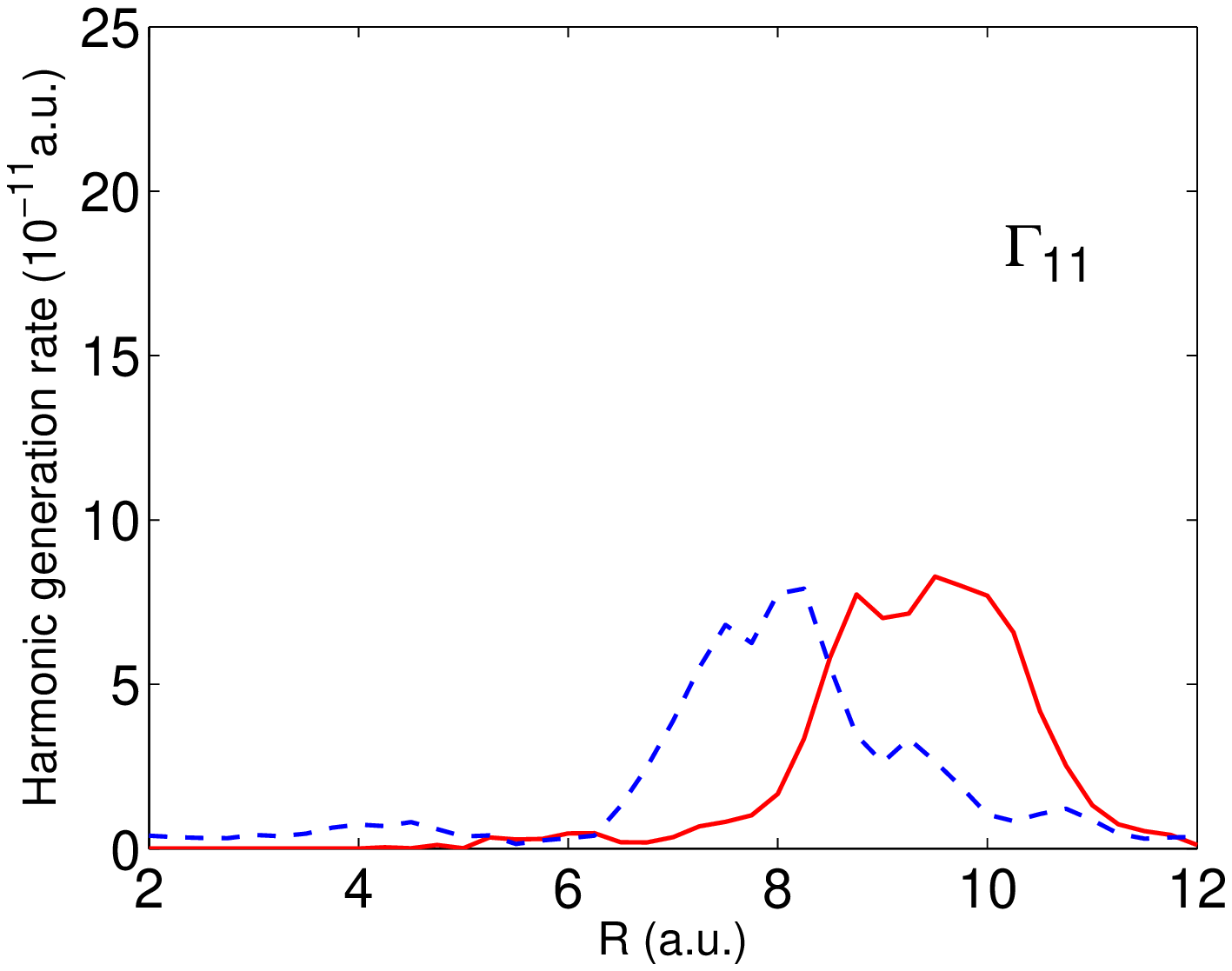}} 
%      \subfigure[]{\includegraphics[width=0.28\textwidth]{HHG_F0p0533_w0p0428_eta0p251_Nt256_Gn7.eps}} 

\caption{Harmonic generation rates as functions of internuclear separation $R$ for the lower (solid red) and upper (dashed blue) of the $\mbox{H}^{+}_{2}$ ion. The harmonic order is shown on each plot. The laser field parameters are $I =1\times 10^{14} W/cm^{2}$ and $\lambda = 1064\,nm$. }  % -0.506673584605708 - 0.009964962098731*I:
\label{fig:1}       % Give a unique label
\end{figure}

In Figure 5 we show the rates $\Gamma_{n}$ as functions of $R$ for the lower (red) and upper (dashed blue) states of the ion using a common linear scale. As can be seen in Fig.~5c,  when $R$ varies from the equilibrium separation $2$ to $12\,au$, the $5$th harmonic for the lower and upper states has very similar large peaks around $R\approx5\,au$, where the first enhancement of the ionization rate (Figure 4a) occurs. Note, however that the ionization rate patterns for the two states are not as similar as those of $\Gamma_{5}$. Observing $\Gamma_{7}$ and $\Gamma_{9}$ in panels (d, e) we find a shift in the peaks towards larger $R$. For the lower state a large peak in the ionization rate appears around $R\approx 9\,au$ (Figure 4a), and $\Gamma_{9}$ also displays a peak there (Figure 5e). Meanwhile, the upper-state $\Gamma_{9}$ rate deviates for $R>8\,au$, somewhat in accord with its decrease in ionization rate. 

The HG rates for higher orders of $n$ (beyond $n=11$) become smaller, thus we do not show them here. Together Figures 4a and 5 demonstrate that the enhancement of the ionization rate for the lower and upper states of the hydrogen molecular ion can be linked to an enhancement of the harmonic generation rates in certain $R$-ranges. This happens for harmonic orders within the molecular plateau region $n< n_{M}$.  

As a further demonstration of the correspondence we show the HG rate as function of $R$ for $\lambda = 800\,nm$ in Figure 6, while the ionization rates are given in Figure 4b. The upper state shows a very prominent ionization peak around $R=8\,au$. 

\begin{figure}[h]
\centering
%  \includegraphics[width=0.55\textwidth]{Ad_diab_en_F0p0533_lambda288.eps}
%  \includegraphics[width=0.50\textwidth]{h2p_ion_rate_1su.pdf}
      \subfigure[]{\includegraphics[width=0.320\textwidth]{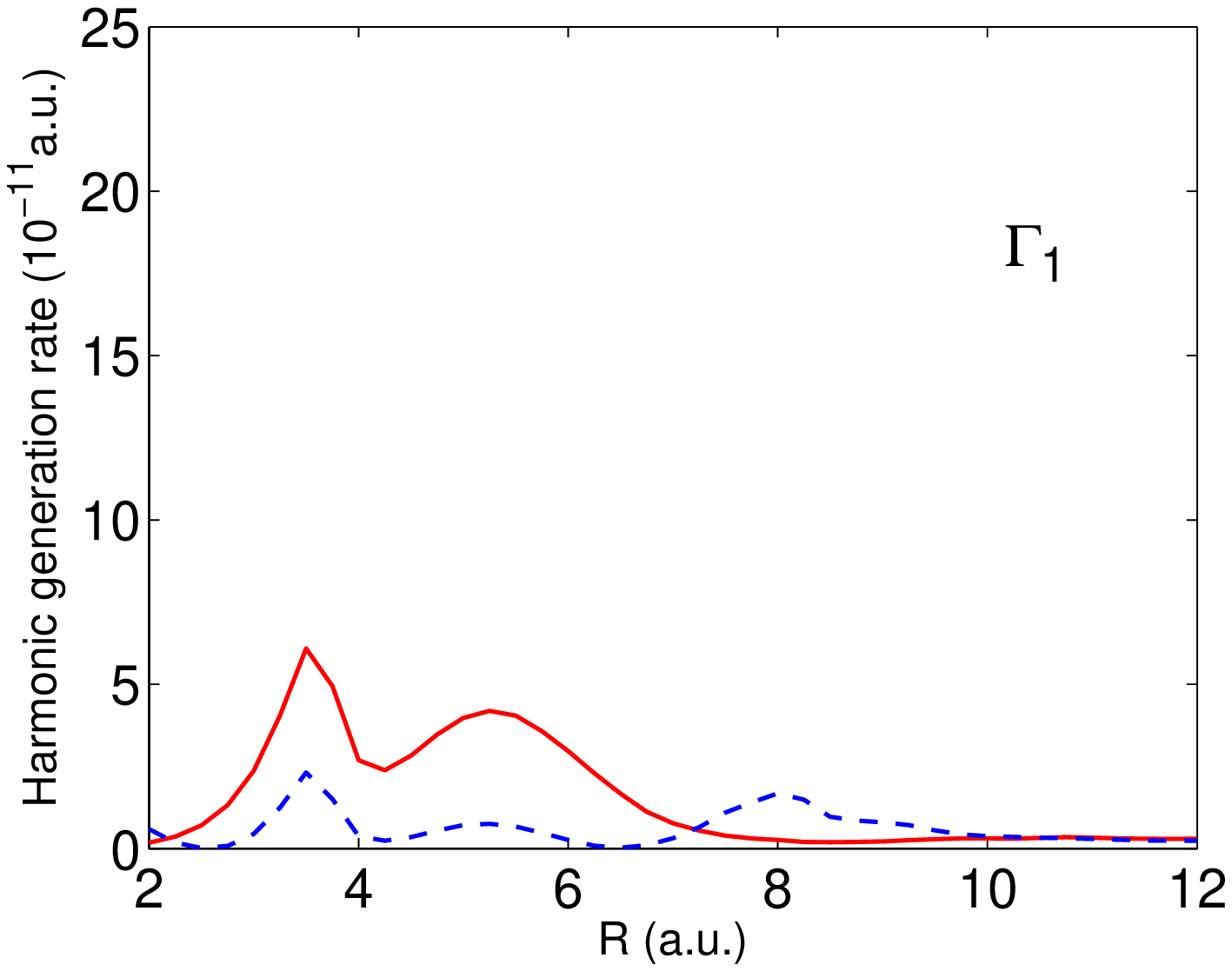}}
      \subfigure[]{\includegraphics[width=0.320\textwidth]{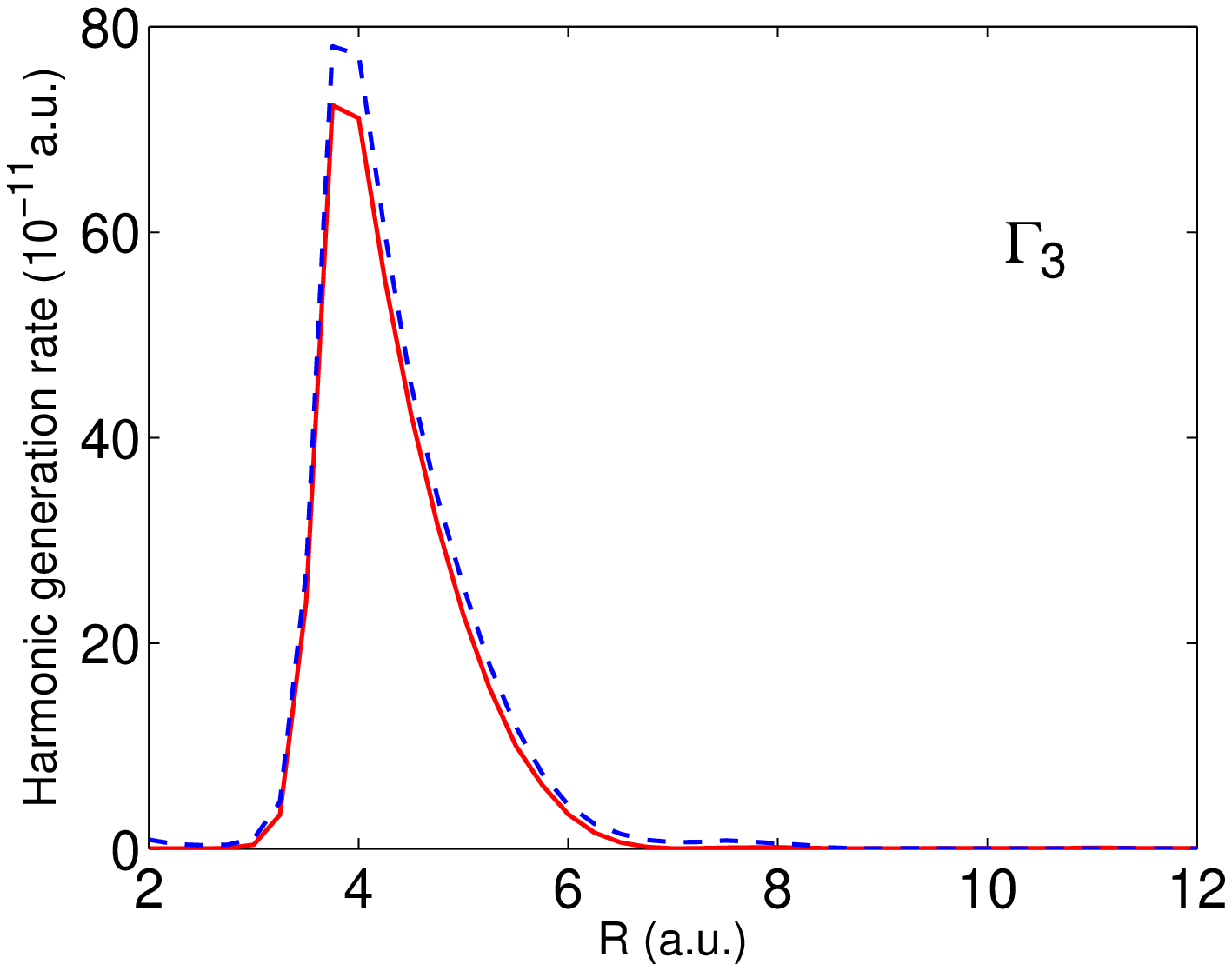}}
      \subfigure[]{\includegraphics[width=0.320\textwidth]{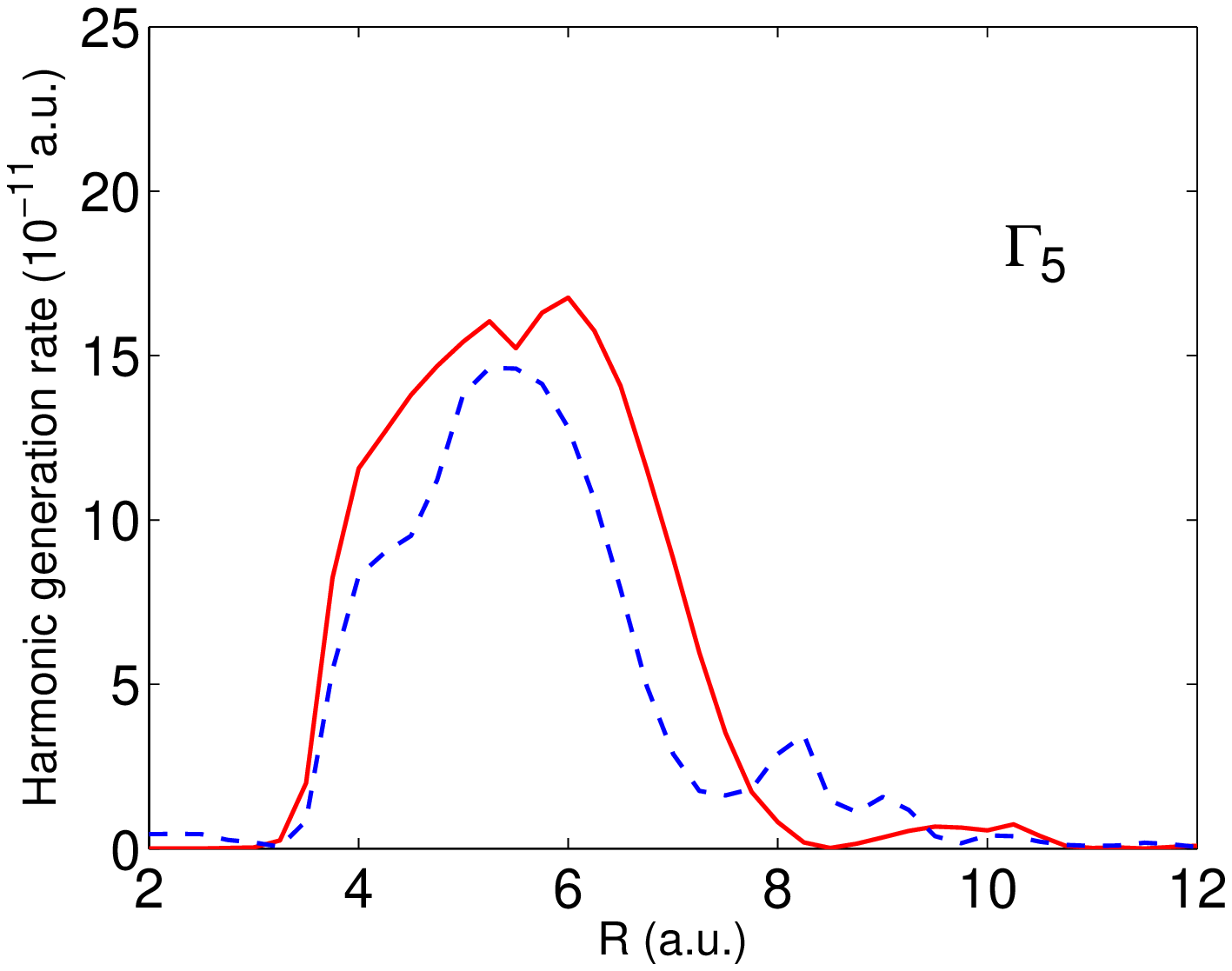}}
      \subfigure[]{\includegraphics[width=0.320\textwidth]{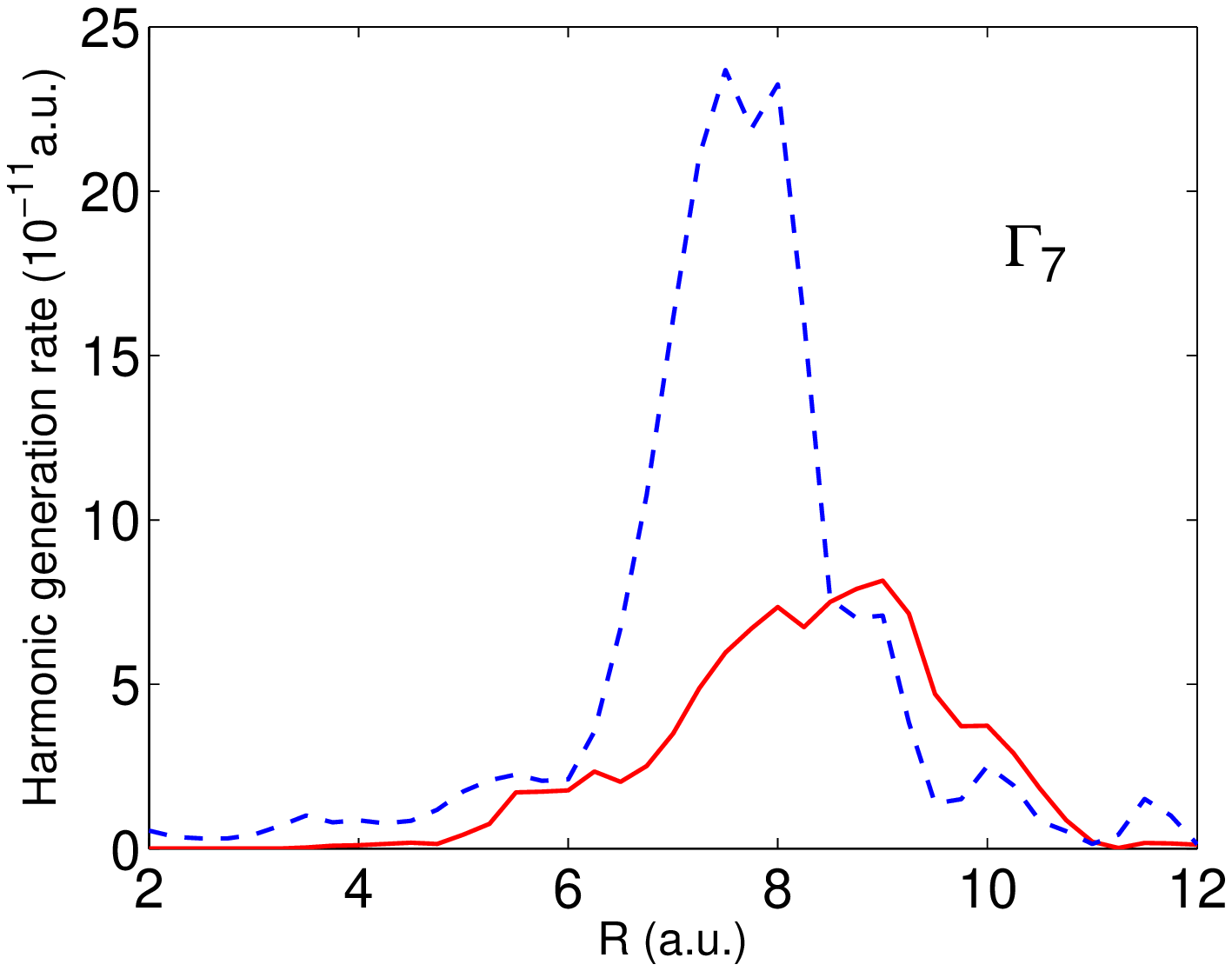}}
      \subfigure[]{\includegraphics[width=0.320\textwidth]{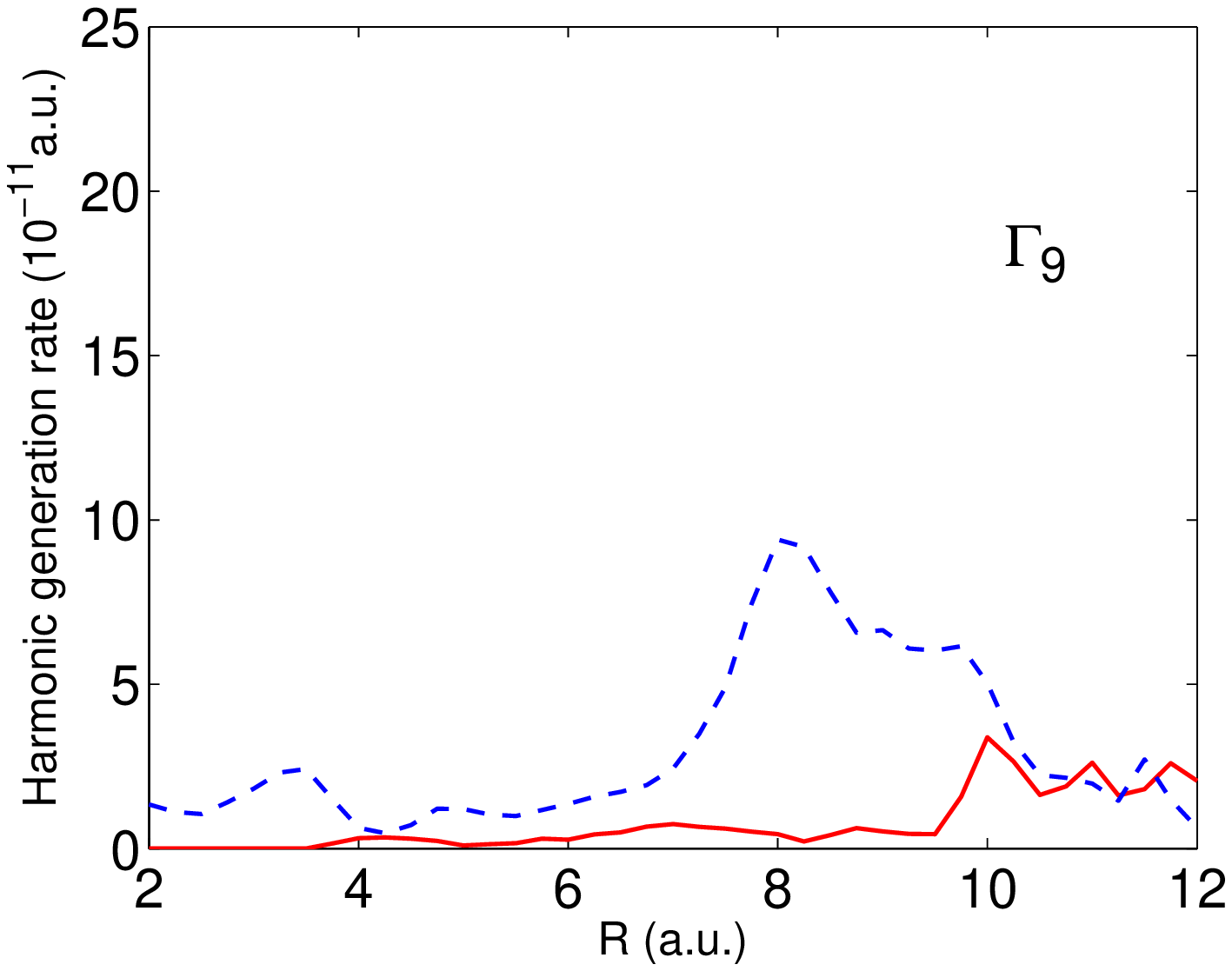}}
      \subfigure[]{\includegraphics[width=0.320\textwidth]{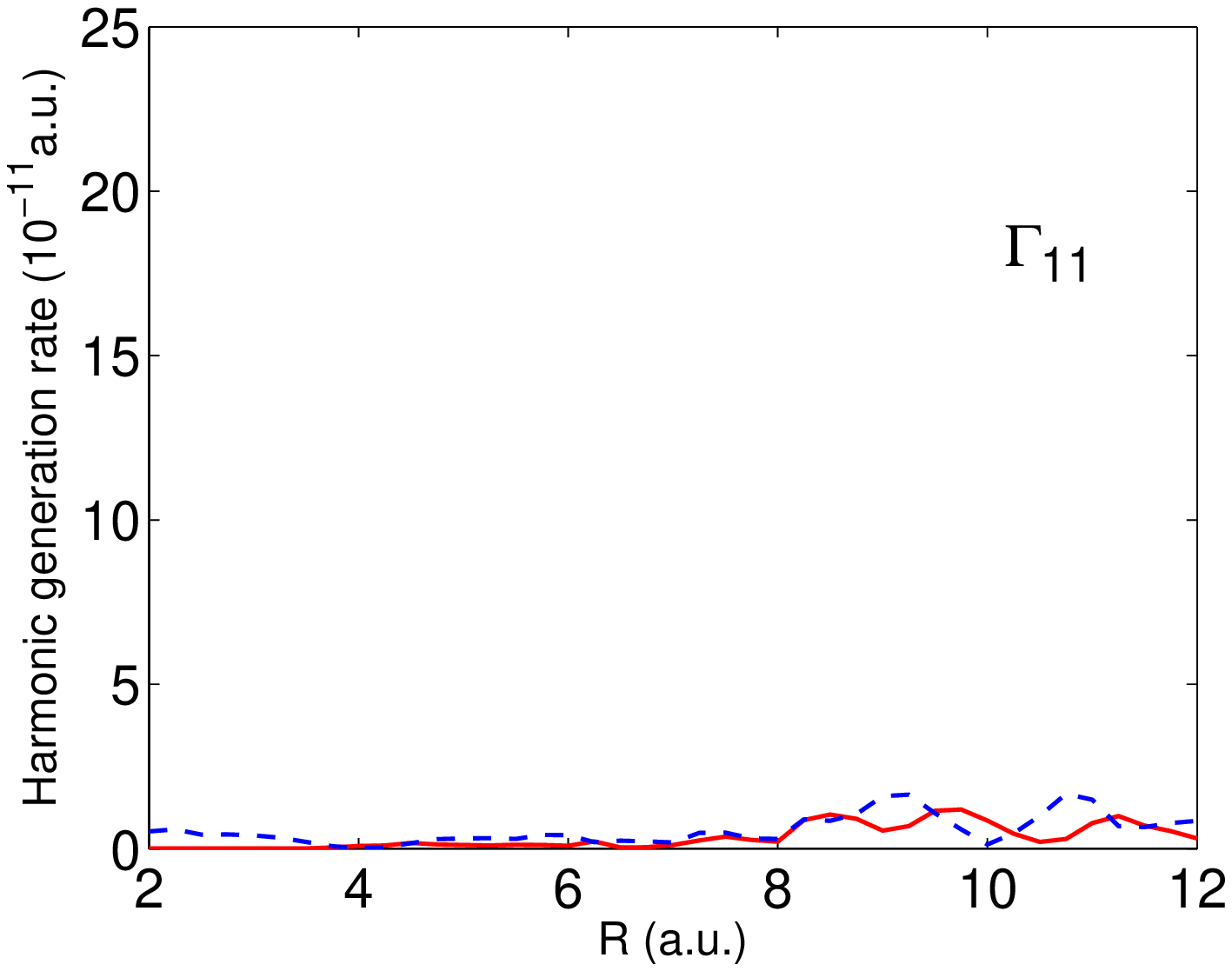}} 
%      \subfigure[]{\includegraphics[width=0.28\textwidth]{HHG_F0p0533_w0p0428_eta0p251_Nt256_Gn7.eps}} 

\caption{ The same plots shown as in Figure 5, however for $\lambda = 800\,nm$.}  % -0.506673584605708 - 0.009964962098731*I:
\label{fig:1}       % Give a unique label
\end{figure}

As can be observed from Figure 6b, $\Gamma_{3}$ displays strong peaks around $R\approx4\,au$  in a region for both states, where the ionization rates are not strong. This is particularly true for the lower state. This behavior, thus, must come from bound-state populations \cite{Zuo2}. However, a major peak can be observed in $\Gamma_{7}$ around $R\approx 8\,au$ (Figure 6d) for the upper state, where a strong enhancement of the ionization rate does occur (Figure 4b). A significant feature around $R\approx 8\,au$ is also visible in $\Gamma_{9}$ for the upper state (Figure 6e).

Compared to the $\lambda = 1064\,au$ case Fig.~6f shows a much weakened $\Gamma_{11}$ rate. This is a consequence of the molecular cutoff law (\ref{nMol}). Naively, one might think that a reduced wavelength $\lambda$ (increased $\omega$) will simply push the cutoff to higher $R$.  However, at $R>10\,au$ the ionization rates drop (cf. Fig.~4).  One reason why the dynamics change at large $R$ is that the upper and lower states become nearly degenerate, i.e., $\omega \gg \epsilon_{1\sigma_{u}} - \epsilon_{1\sigma_{g}} $, and the strong-coupling regime \cite{Zuo2} is reached. 

From the behavior of the $n<n_{M}$ rates $\Gamma_{n}$ shown in Figs.~5, 6 it is obvious that the dipole moment  Eq.(\ref{dipmom}) is a complicated periodic function of time when $R\gg 2\,au$. It is no longer dominated by the fundamental frequency $\omega$ as is usual for atomic HG spectra.

\section{Conclusions}

We have presented a non-Hermitian Floquet calculation of HHG for the lowest two $\mbox{H}^{+}_{2}$ eigenstates in monochromatic strong laser fields using the length gauge. A pseudospectral representation of the Hamiltonian was applied, and the CAP method was implemented to avoid the calculation of an oscillatory tail in the coupled-channel resonance wave function and to get accurate resonance parameters. In this approach even though we needed to solve a large non-Hermitian matrix problem to get the solution for the TDSE, we avoided problems that are associated with time-stepping algorithms, particularly the accumulation of phase errors.    

The results for the HHG rates for the lower state of $\mbox{H}^{+}_{2}$ in strong laser fields were compared with previous calculations in the literature \cite{TelChu2, Zuo2}. The cutoff positions in the HHG spectra were examined and compared with the classical cutoff formula given in \cite{Lew}. Good agreement was found for $I \leq10^{14} W/cm^{2}$ and some deviations at higher intensities for high harmonic orders.

In the region of the molecular plateau, $n<n_{M}$, (cf. Eq.(\ref{nMol})), while considering separations $R=4-10\,au$ it is shown that the enhancement of the ionization rate for the lower and upper states at $R\approx  8\,au$ causes an enhancement of the harmonic generation rate in the following way: $\Gamma_{5}$ for both states and $\Gamma_{9}$ for the lower state of $\mbox{H}^{+}_{2}$ in a field of $I =10^{14} W/cm^{2}$ and $\lambda = 1064\,nm$ show peaks; similarly $\Gamma_{7}$ for the upper state of $\mbox{H}^{+}_{2}$ in a field of $I =10^{14} W/cm^{2}$ and $\lambda = 800\,nm$ displays peaks at $R_{c}\approx 8\,au$. The lower bound of this enhanced HG region $R=4\,au$ is characterized by near-resonant coupling $\omega\approx \epsilon_{1\sigma_{u}}- \epsilon_{1\sigma_{g}}$. It is bounded at $R\approx 10\,au$ by the decrease in ionization rate (cf. Fig.~4). On the other hand, for $\lambda = 800\,nm$ $\Gamma_{3}$ dominates the HG spectra around $R\approx 4\,au$ for the upper and lower states which cannot be associated with an enhanced ionization rate, but with strong resonant coupling, as explained in Ref.\cite{Zuo2}. 

% in field of $I =10^{14} W/cm^{2},\lambda = 1064\,nm$
%in field of $I =10^{14} W/cm^{2}$ and $\lambda = 800\,nm$

%for the two states at certain internuclear separations where the ionization rates peak.  

\section*{Acknowledgments}
We thank the Natural Sciences and Engineering Research Council
of Canada for financial support, and the Shared Hierarchical Academic Research Computing Network \cite{SHA}
for access to high-performace computing facilities. We thank Mitsuko Murakami for helpful discussions. 
%\newpage


\begin{thebibliography}{99}


% basic HHG
\bibitem{Piraux} See {\it Proceedings of the Workshop, Super Intense Laser-Atom Physics (SILAP) III}, edited by B.~Piraux et al (1993).
\bibitem{Kul} K.~C.~Kulander, K.~J.~Schafer and L.~J.~Krause, in {\it Proceedings of the Workshop, Super Intense Laser-Atom Physics (SILAP) III}, (\cite{Piraux}). 
\bibitem{Cor} P.~B.~Corkum, Phys.\ Rev.\ Lett.\ \textbf{71}, 1994 (1993)
\bibitem{Lew} M.~Lewenstein, Ph.~Balcou, M.~Yu.~Ivanov, A.~L'Huillier and P.~.B.~Corkum, Phys. Rev. A. \textbf{49}, 2117 (1994)
\bibitem{Kop} R.~Kopold, W.~Becker, and M.~Kleber, Phys. Rev. A. \textbf{58}, 4022 (1998)
%
\bibitem{Krause} L.~J.~Krause, K.~J.~Schafer and K.~C.~Kulander, Chem. Phys. Lett. \textbf{178}, 573 (1991).
\bibitem{Zuo1} T.~Zuo, S.~Chelkowski, and A.~D.~Bandrauk, Phys. Rev. A. \textbf{48}, 3837 (1993). 
\bibitem{Zuo2} T.~Zuo, S.~Chelkowski, and A.~D.~Bandrauk, Phys. Rev. A. \textbf{49}, 3943 (1993). 

% Floquet HHG
\bibitem{ShakePot} R.~M.~Potvliege and R.~Shakeshaft, Phys.\ Rev.\ A \textbf{40}, 3061 (1989)
\bibitem{Burke} P.~G.~Burke, P.~Francken, and C.~J.~Joachain, EuroPhys. Lett. \textbf{13}, 617 (1990)
\bibitem{Burke2} P.~G.~Burke, P.~Francken, and C.~J.~Joachain, J.~Phys. B: \textit{At. Mol. Opt. Phys.} \textbf{24}, 761 (1991)
\bibitem{Ben} N.~Ben-Tal, N.~Moiseyev, C.~Leforestier and R.~Kosloff,  J.~.Chem. Phys. \textbf{94}, 7311 (1991)
\bibitem{Ben2} N.~Ben-Tal, N.~Moiseyev, R.~Kosloff and C.~Cerjan,  J.~Phys. B: \textit{At. Mol. Opt. Phys.} \textbf{26}, 1445 (1993)

% Floquet HHG by Chu and his co-workers
\bibitem{TelChu1} D.~A.~Telnov, J.~Wang and S.~I.~Chu, Phys.\ Rev.\ A \textbf{52}, 3988 (1995)
\bibitem{TongChu} X.~M.~Tong and S.~I.~Chu, Chem. Phys. \textbf{217}, 119 (1996)
\bibitem{XiChu} X.~Chu and S.~I.~Chu, Phys.\ Rev.\ A \textbf{63}, 023411 (2001)
\bibitem{TelChu2} D.~A.~Telnov, and S.~I.~Chu, Phys.\ Rev.\ A \textbf{71}, 013408 (2005)
\bibitem{TelChu3} D.~A.~Telnov, and S.~I.~Chu, Phys.\ Rev.\ A \textbf{76}, 043412 (2007)
% our
\bibitem{TsogMarko2} Ts.~Tsogbayar and M.~Horbatsch, J. Phys. B \textbf{46}, 46, 245005 (2013)  

% Floq th
\bibitem{Floq} G.~Floquet, Ann.\ Ec.\ Norm. \ Suppl. \textbf{12}, 47 (1883) 
\bibitem{Sambe} H.~Sambe, Phys.\ Rev.\ A \textbf{7}, 2203 (1973)
% Floq th
% H2+
\bibitem{Hyl} E. A.~Hylleraas, Z.\ Physik, \textbf{71}, 739 (1931) 
\bibitem{Jaffe} G.~Jaff\'e, Z.\ Physik, \textbf{87}, 535 (1934) 
\bibitem{Bab} W. G.~Baber and H. R.~ Hass\'e, Proc.\ Camb.\ Phil.\ Soc. \textbf{31}, 564 (1935) 
% H2+
% Electrodynamics
\bibitem{Land} L.~D.~Landau and E.~.M.~Lifshitz, \textit{The classical theory of fields}, Pergamon-Oxford, (1975)
% Electrodynamics
% Spect method
%\bibitem{Fun} D.~Funaro, \textit{Polynomial Approximation of Differential Equations}, Springer-Verlag, (1992)
%\bibitem{Hest} J.S.~Hesthaven, S.~Gottlieb and D.~Gottlieb, \textit{Spectral Methods for Time-Dependent Problems}, Cambridge University Press, (2007)
\bibitem{TsogMarko1} Ts.~Tsogbayar and M.~Horbatsch, J. Phys. B \textbf{46}, 46, 085004 (2013)  
\bibitem{Xichu2} Xi Chu and Shih-I Chu, Phys.\ Rev.\ A \textbf{63}, 013414 (2000)
% Spect method
% CAP
%\bibitem{Riss} U.V.~Riss and H.-D.~Meyer, J.~Phys.\ B\ \textbf{26}, 4503 (1993)
%\bibitem{Tsog} Ts.~Tsogbayar and M.~ Horbatsch Few-Body Syst. \textbf{54}, 431 (2013)
%\bibitem{Lef} R.~Lefebvre, M.~Sindelka and N.~Moiseyev, Phys.\ Rev.\ A\ \textbf{72}, 052704 (2005)
%\bibitem{Santra} R.~Santra, Phys.\ Rev.\ A\ \textbf{74}, 034701 (2006) 
%\bibitem{Ackad} E.~Ackad and M.~Horbatsch, Phys.\ Rev.\ A\ \textbf{76}, 022503 (2007)
% CAP
% Electrodynamics
%\bibitem{Land} L.~D.~Landau and E.~.M.~Lifshitz, \textit{The classical theory of fields}, Pergamon-Oxford, (1975)
% Electrodynamics
%%% Enhancement of ionization rate
\bibitem{Zuob} T.~Zuo and A. D.~Bandrauk, Phys.\ Rev.\ A \textbf{52}, R2511 (1995)
\bibitem{Plum} M.~Plummer and J. F.~McCann, J.\ Phys.\ B \textbf{29}, 4625 (1996) 
\bibitem{Mul} Z.~Mulyukov, M.~Pont, and R.~Shakeshaft, Phys.\ Rev.\ A \textbf{54}, 4299 (1996)
\bibitem{Mad} L.~B.~Madsen and M.~Plummer, J.\ Phys.\ B \textbf{31}, 87 (1998) 

%%% Enhancement of ionization rate



% exp HHG
%\bibitem{Codf} K.~Codling and L. J.~Frasinski, J.\ Phys.\ B \textbf{26}, 783 (1993)
%\bibitem{Gius} A.~Giusti-Suzor, $et$ $al.,$ J.\ Phys.\ B \textbf{28}, 309 (1995) 
%\bibitem{Walsh} T.~D.~G.~Walsh, L.~Strach and S.~L.~Chin, J.\ Phys.\ B \textbf{31}, 4853 (1998) 
%\bibitem{Ben} I.~Ben-Itzhak $et$ $al.,$ Phys.\ Rev.\ A \textbf{78}, 063419 (2008) %PRA 78, 063419 (2008)
%\bibitem{Huang} Y.~Huang and Shih-I Chu Chem. \ Phys.\ Lett. \textbf{225}, 46 (1994)
% \bibitem{Xichu} Xi Chu and Shih-I Chu, Phys.\ Rev.\ A \textbf{63}, 013414 (2000)
% \bibitem{Smir} O.~Smirnova $et$ $al$, Nature \textbf{460}, 972 (2009)
% \bibitem{Huang} Y.~Huang and Shih-I Chu Chem. \ Phys.\ Lett. \textbf{225}, 46 (1994)
%\bibitem{Ivanov} M.~Ivanov, T.~Seideman, P.~Corkum, F.~Ilkov and P.~Dietrich, Phys.\ Rev.\ A \textbf{54}, 1541 (1995) 
%\bibitem{Chel} S.~Chelkowski and A.~D.~Bandrauk  J. Phys. B \textbf{28}, L723 (1995)
%\bibitem{Gius2} A.~Giusti-Suzor and F.~ H.~Mies, Phys.\ Rev.\ Lett.\ \textbf{68}, 3869 (1992) 
 %\bibitem{Trump} C.~Trump $et$ $al.,$  Phys.\ Rev.\ A \textbf{62}, 063402 (2008) % PRA 62, 063402 (2000)
%\bibitem{Pav} D.~Pavi\v{c}i\'c, A.~Kiess, T.~ W.~H\"{a}nsch, and H.~Figger, Phys.\ Rev.\ Lett.\ \textbf{94}, 163002 (2005) 
%PRA 62, 063402 (2000)
% exp diatomic molecule
% theor rs
%\bibitem{Plum} M.~Plummer and J. F.~McCann, J.\ Phys.\ B \textbf{29}, 4625 (1996) 
%\bibitem{Mul} Z.~Mulyukov, M.~Pont, and R.~Shakeshaft, Phys.\ Rev.\ A \textbf{54}, 4299 (1996)
%\bibitem{Xichu} Xi Chu and Shih-I Chu, Phys.\ Rev.\ A \textbf{63}, 013414 (2000)
%\bibitem{TsogMarko} Ts.~Tsogbayar and M.~Horbatsch, J. Phys. B \textbf{46}, 46, 085004 (2013)  
%\bibitem{Seid} T.~Seideman, M.~Yu.~Ivanov and P.~B.~Corkum , Phys.\ Rev.\ Lett.\ \textbf{75}, 2819 (1995) 
%\bibitem{ShakePot} R.~Shakeshaft and R.~M.~Potvliege, Phys.\ Rev.\ A \textbf{36}, 5478 (1987)
%\bibitem{Mad} L.~B.~Madsen and M.~Plummer, J.\ Phys.\ B \textbf{31}, 87 (1998) 
% theor rs
% Floq th
%\bibitem{Floq} G.~Floquet, Ann.\ Ec.\ Norm. \ Suppl. \textbf{12}, 47 (1883) 
%\bibitem{Sambe} H.~Sambe, Phys.\ Rev.\ A \textbf{7}, 2203 (1973)
% Floq th
% H2+
%\bibitem{Hyl} E. A.~Hylleraas, Z.\ Physik, \textbf{71}, 739 (1931) 
%\bibitem{Jaffe} G.~Jaff\'e, Z.\ Physik, \textbf{87}, 535 (1934) 
%\bibitem{Bab} W. G.~Baber and H. R.~ Hass\'e, Proc.\ Camb.\ Phil.\ Soc. \textbf{31}, 564 (1935) 
% H2+
% Spect method
%\bibitem{Fun} D.~Funaro, \textit{Polynomial Approximation of Differential Equations}, Springer-Verlag, (1992)
%\bibitem{Hest} J.S.~Hesthaven, S.~Gottlieb and D.~Gottlieb, \textit{Spectral Methods for Time-Dependent Problems}, Cambridge University Press, (2007)
%\bibitem{Kos} V.~Kokoouline, O.~Dulieu, R.~Kosloff and F.~Masnou-Seeuws, J.\ Chem.\ Phys. \textbf{110}, 9865 (1999)
%\bibitem{Tel} D.~A.~Telnov and Shih-I Chu, Phys.\ Rev.\ A \textbf{71}, 013408 (2005)
%\bibitem{Tao} L.~Tao, C.~W.~McCurd and T.~N.~Rescigno, Phys.\ Rev.\ A \textbf{80}, 013402 (2009)
%\bibitem{Guan} X.~Guan, E.~B.~Secor, K.~Bartschat and B.~I.~Schneider, Phys.\ Rev.\ A \textbf{84}, 033420 (2011)
%\bibitem{Guan2} X.~Guan, R.~C.~DuToit and K.~Bartschat, Phys.\ Rev.\ A \textbf{87}, 053410 (2013)
% Spect method
% CAP
%\bibitem{Riss} U.V.~Riss and H.-D.~Meyer, J.~Phys.\ B\ \textbf{26}, 4503 (1993)
%\bibitem{Tsog} Ts.~Tsogbayar and M.~ Horbatsch ... (2011)
%\bibitem{Lef} R.~Lefebvre, M.~Sindelka and N.~Moiseyev, Phys.\ Rev.\ A\ \textbf{72}, 052704 (2005)
%\bibitem{Santra} R.~Santra, Phys.\ Rev.\ A\ \textbf{74}, 034701 (2006) 
%\bibitem{Ackad} E.~Ackad and M.~Horbatsch, Phys.\ Rev.\ A\ \textbf{76}, 022503 (2007)
% CAP
%\bibitem{Zuob} T.~Zuo and A. D.~Bandrauk, Phys.\ Rev.\ A \textbf{52}, R2511 (1995)
%\bibitem{Kos} V.~Kokoouline, O.~Dulieu, R.~Kosloff and F.~Masnou-Seeuws, J.\ Chem.\ Phys. \textbf{110}, 9865 (1999)
%
\bibitem{SHA} SHARCNET: Shared Hierarchical Academic Research Computing Network. http://www.sharcnet.ca
%

%%% DC

% exp diatomic molecule
%\bibitem{Codf} K.~Codling and L. J.~Frasinski, J.\ Phys.\ B \textbf{26}, 783 (1993)
%\bibitem{Gius} A.~Giust-Suzor, $et$ $al.,$ J.\ Phys.\ B \textbf{28}, 309 (1995) 
%\bibitem{Post} J. H.~Posthumus $et$ $al.,$ J.\ Phys.\ B \textbf{29}, L525 (1996) 
%\bibitem{Cons} E.~Constant, H.~Stapelfeldt, and P. B.~Corkum, Phys.\ Rev.\ Lett.\ \textbf{76}, 4140 (1996) 
%\bibitem{Gib} G. N.~Gibson, M.~Li, C.~Guo, and J.~Neira, Phys.\ Rev.\ Lett.\ \textbf{79}, 2022 (1997) 
%\bibitem{Stap} H.~Stapelfeldt $et$ $al.,$ Phys.\ Rev.\ A \textbf{58}, 426 (1998)
%\bibitem{Will} I. D.~Williams, $et$ $al.,$ J.\ Phys.\ B \textbf{33}, 2743 (2000) 
%\bibitem{Erg} Th.~Ergler, $et$ $al.,$ Phys.\ Rev.\ Lett.\ \textbf{95}, 093001 (2005) 
%\bibitem{Ben} I.~Ben-Itzhak $et$ $al.,$ Phys.\ Rev.\ A \textbf{78}, 063419 (2008)
% exp diatomic molecule
% theor rs
%\bibitem{Zuob} T.~Zuo and A. D.~Bandrauk, Phys.\ Rev.\ A \textbf{52}, R2511 (1995)
%\bibitem{Liang} Liang-You Peng, $et$ $al.,$ J.~Chem.\ Phys.\ \textbf{120}, 10046 (2004)
%\bibitem{Kel} L. V.~Keldysh, Zh.~\'Eksp.\ Teor.\ Fiz .\ \textbf{47}, 1945 (1964) [Sov.\ Phys.\ JETP \textbf{20}, 1307 (1965)]
%\bibitem{Post95} J. H.~Posthumus $et$ $al.,$ J.\ Phys.\ B \textbf{28}, L349 (1995)
%\bibitem{Band}  A. D.~Bandrauk and H.Z.~Lu, Phys.\ Rev.\ A \textbf{62}, 053406 (2000)
% theor rs
% H2+ exp. dissoc. & ioniz
%\bibitem{Buck} P. H.~ Bucksbaum, A.~Zavriyev, H. G.~Muller, and D. W.~Schumacher, \ Phys.\ Rev.\ Lett.\ \textbf{64}, 1883 (1990) 
% H2+ exp. dissoc. & ioniz
% H2+ solution
%\bibitem{Tsog2} Ts.Tsogbayar,  J.\ Phys.\ B \textbf{42}, 165007 (2009) 
% H2+ solution
% H atom solution
%\bibitem{TsogMarko2} Ts.~Tsogbayar and M.~Horbatsch, Few-Body Syst  \textbf{54}, 431 (2013)
%\bibitem{Kosl} E.~Fattal, R.~Baer and R.~Kosloff, Phys.\ Rev.\ E \textbf{53}, 1217 (1996)
% H atom solution
% exp
%\bibitem{Sta} A.~Staudte $et$ $al.,$ Phys.\ Rev.\ Lett. \textbf{98}, 073003 (2007)
%
%\bibitem{SHA} SHARCNET: Shared Hierarchical Academic Research Computing Network. http://www.sharcnet.ca
%

\end{thebibliography}
\end{document}